\documentclass[12pt,a4paper]{article}
\usepackage[latin9]{inputenc}
\usepackage{a4wide}
\usepackage{slashed}
\usepackage{epsf}
\usepackage{amssymb}
\usepackage{ifpdf}
\usepackage[notref,notcite,final]{showkeys}
\ifpdf
\usepackage[pdftex]{graphicx}
\usepackage[pdftex,unicode,implicit]{hyperref}
\hypersetup{%
  pdftitle    = {¡Aún más guarradas con N=2!},
  pdfkeywords = {N=2 Supersymmetry, BPS solutions, vector and hyper multiplets},
  pdfauthor   = {M. Hübscher, P. Meessen, T. Ortín},
  pdfcreator  = {pdf\LaTeXe\ with package \flqq hyperref\frqq},
  pdfproducer = {pdf\LaTeXe\ with package \flqq hyperref\frqq},
  pdfpagemode = None,  
  pdffitwindow= true,  
  unicode     = true,
  plainpages  = false,
  colorlinks  = true,  
  citecolor   = black,
  urlcolor    = black,
  linkcolor   = black
}
\newcommand{\hepth}[1]{{\tt \href{http://www.arXiv.org/abs/hep-th/#1}{hep-th/#1}}}

\else
  \usepackage[dvips]{graphicx}
  \usepackage[unicode,implicit]{hyperref}
  \newcommand{\hepth}[1]{{\tt hep-th/#1}}

\fi

\makeatletter
\@addtoreset{equation}{section}
\makeatother

\begin{document}

\begin{flushright}
\small
IFT-UAM/CSIC-06-32\\
{\bf hep-th/0606281}\\
June 30, 2006
\normalsize
\end{flushright}

\begin{center}

\vspace{2cm}

{\Large\bf 
  Supersymmetric solutions of $N=2$ $D=4$ SUGRA:\\[.4cm] 
  \ifpdf
    \href{http://en.wikipedia.org/wiki/Shebang}{the whole ungauged shebang}
  \else
    the whole ungauged shebang
  \fi
}

\vspace{2cm}

\ifpdf
{\bf \href{mailto:mechthild.huebscher@uam.es}{Mechthild H\"ubscher}},
{\bf \href{mailto:Patrick.Meessen@cern.ch}{Patrick Meessen}}, and 
{\bf \href{mailto:Tomas.Ortin@cern.ch}{Tom\'as Ort\'{\i}n}}\\[.4cm]
\else
{\bf Mechthild H\"ubscher},
{\bf Patrick Meessen}, and 
{\bf Tom\'as Ort\'{\i}n}\\[.4cm]
\fi

{\em 
\ifpdf 
   \href{http://gesalerico.ft.uam.es/}{Instituto de F\'{\i}sica Te\'orica UAM/CSIC}
\else
   Instituto de F\'{\i}sica Te\'orica UAM/CSIC
\fi
, Facultad de Ciencias, C-XVI\\
C.U. Cantoblanco, E-28049 Madrid, Spain.
}

\vspace{2cm}


{\bf Abstract}

\end{center}

\begin{quotation}
  
  \small In this article we complete the classification of the supersymmetric
  solutions of $N=2$ $D=4$ ungauged supergravity coupled to an arbitrary
  number of vector- and hypermultiplets. 
  
  We find that in the timelike case the hypermultiplets cause the
  constant-time hypersurfaces to be curved and have $SU(2)$ holonomy identical
  to that of the hyperscalar manifold. The solutions have the same
  structure as without hypermultiplets but now depend on functions which are
  harmonic in the curved 3-dimensional space. We discuss an example obtained
  from a hyper-less solution via the c-map.
  
  In the null case we find that the hyperscalars can only depend on the null
  coordinate and the solutions are essentially those of the hyper-less case.

\end{quotation}

\newpage

\pagestyle{plain}


\tableofcontents

\newpage

\section{Introduction}

The classification of all the supersymmetric configurations of $N=2,d=4$
ungauged supergravity coupled to vector multiplets has recently been achieved
in Ref.~\cite{Meessen:2006tu} and it is only natural to try to extend those
results to more general couplings of $N=2,d=4$ supergravities
\cite{deWit:1984pk,deWit:1984px} since, after all, generic Calabi-Yau
compactifications yield theories with more than just vector multiplets. 
The simplest extension, which just happens to be the one we are going to 
consider in this paper, is the inclusion in the
theory of an arbitrary number of hypermultiplets. 

This is a problem that has, so far, largely been ignored in the literature on the
grounds that hypermultiplets do not couple to the vector multiplets at low
energies and, therefore, their presence was irrelevant to study, for instance,
black-hole solutions.\footnote{
 See, however, Ref.~\cite{Celi:2003qk} and references therein.
}
Their generic presence is, however, to be expected, and,
in general, hypermultiplets will be excited and their non-triviality will
certainly modify the known solutions since they couple to gravity.

What are we to expect? In order to answer this question it is worthwhile to
have a look at the c-map of the general {\em cosmic string} solution found in
Ref.~\cite[(5.93)]{Meessen:2006tu}:

\begin{equation}
  \label{eq:CosmString}
  \begin{array}{rclrcl}
    ds^{2} & =& 2\ du\ dv \; -\; 2e^{-\mathcal{K}}\ dz\ dz^{*}\, ,\hspace{1cm}
    & Z^{i} & =& Z^{i}(z) \, , \\
      & & & &  & \\
    F^{\Lambda} & = &  0\, ,   & q^{u} & = & {\rm const.} \, ,\\
  \end{array}
\end{equation}

This solution is especially suited for our purposes since it has an extremely
simple form, is 1/2-BPS, and the corresponding Killing spinor is constant,
thus ensuring that the dual solution is at least 1/2-BPS.  Using the formulae
in Appendix~\ref{appsec:dualquat} we can dualize the above solution along the
spacelike direction $u-v$, to another solution in minimal supergravity coupled
to a certain number of hypermultiplets; the resulting spacetime metric is the
one above, the graviphoton field strength still vanishes, and some of the
hyperscalars have a (anti-)holomorphic spacetime dependency (the details are
spelled out in Sec.~\ref{sec:CosmString}.).  Comparing this to the general
timelike solution in Tod's classification of supersymmetric solutions in
minimal $N=2$ $d=4$ supergravity \cite{Tod:1983pm}, we reach the conclusion
that having non-trivial hyperscalars must lead to a non-trivial metric on the
constant-time hypersurfaces.
\par
The reason for this occurrence is to be found in the Killing spinor: the relevant 
gravitino variation equation for vanishing graviphoton field strength reads 
schematically $\mathfrak{D}\epsilon \ =\ 0$, where $\mathfrak{D}$ not only contains
the spin connection but also an $\mathfrak{su}(2)$ connection which is constructed
out of hyperscalars. Therefore, if we want BPS solutions with non-trivial scalars,
we need a non-trivial spin connection in order to attain $Hol(\mathfrak{D})=0$, or
said differently: we need to embed one connection into the other.
\par
The embedding of the gauge connection into the spin connection (or the other
way around) was proposed originally in Refs.~\cite{kn:Wilczek,Charap:1977ww}
and used to achieve anomaly cancellation or absence of higher-order
corrections in the context of the Heterotic String in
Refs.~\cite{Candelas:1985en,Khuri:1992hk,Gauntlett:1992nn,Duff:1993yb,Kallosh:1994wy}.
As we are going to see, in this case this mechanism leads to unbroken
supersymmetry through an exact cancellation of the $SU(2)$ and spin
connections in the gravitino supersymmetry transformation, generalizing the
cancellation between $U(1)$ gauge and 2-dimensional spin connection used in
Ref.~\cite{Maldacena:2000mw}.
\par
This embedding turns out to be possible in the timelike case, but not in the
null case; further, it is only in the timelike case that the presence of excited
hyperscalars has important consequences.
\par
Let us summarize our results:
\par
\begin{enumerate}
\item In the timelike case supersymmetric the configurations are completely
  determined by 

  \begin{enumerate}
  \item A 3-dimensional space metric 

    \begin{equation}
    \gamma_{\underline{m}\underline{n}}dx^{m}dx^{n}\, ,\,\,\,\,\, m,n=1,2,3\, ,  
    \end{equation}
    
\noindent
and a mapping $q^{u}(x)$ from it to the quaternionic hyperscalar manifold such that
the 3-dimensional spin connection\footnote{In this paper we use $x,y,z=1,2,3$
  as tangent-space indices or as $SU(2)$ indices.} $\varpi_{x}{}^{y}$ is
related to the pullback of the quaternionic $SU(2)$ connection
$\mathsf{A}^{x}$ by

\begin{equation}
\varpi_{\underline{m}}{}^{xy} = \varepsilon^{xyz}
\mathsf{A}^{z}{}_{u}\ \partial_{\underline{m}}q^{u}\, ,
\end{equation}

\noindent
and such that 

\begin{equation}
\mathsf{U}^{\alpha J}{}_{x}\ (\sigma_{x})_{J}{}^{I} \; =\; 0\, ,
\hspace{1cm}
\mathsf{U}^{\alpha J}{}_{x}\ \equiv\ V_{x}{}^{\underline{m}}\partial_{\underline{m}}q^{u}\
\mathsf{U}^{\alpha J}{}_{u}\, ,
\end{equation}

\noindent
where $\mathsf{U}^{\alpha I}{}_{u}$ is the \textit{Quadbein} defined in
Appendix~\ref{sec-QKG}.

\item A choice of a symplectic vector $\mathcal{I}\equiv \Im{\rm
    m}(\mathcal{V}/X)$ whose components are real harmonic functions with
  respect to the above 3-dimensional metric: 

  \begin{equation}
  \nabla_{\underline{m}}\partial^{\underline{m}}\mathcal{I} =0\, .  
  \end{equation}

  \end{enumerate}
  
  Given $\mathcal{I}$, $\mathcal{R}\equiv \Re{\rm e}(\mathcal{V}/X)$ can in
  principle be found by solving the generalized stabilization equations and
  then the metric is given by 

\begin{equation}
ds^{2} = |M|^{2}(dt+\omega)^{2} 
-|M|^{-2}\gamma_{\underline{m}\underline{n}}dx^{m}dx^{n}\, ,
\end{equation}

\noindent
where

\begin{eqnarray}
|M|^{-2} & = & \langle \mathcal{R} \mid \mathcal{I} \rangle\, ,  \\
& & \nonumber \\
(d\omega)_{xy} & = & 2 \epsilon_{xyz}
\langle\,\mathcal{I}\mid \partial^{z}\mathcal{I}\, \rangle\, .
\end{eqnarray}

The second equation implicitly contains the Dreibein of the 3-dimensional
metric $\gamma$ and its integrability condition is 

\begin{equation}
 \langle \mathcal{I} \mid
 \nabla_{\underline{m}}\partial^{\underline{m}}\mathcal{I} \rangle=0\, .
\end{equation}

As is discussed in {\em e.g.\/}
Refs.~\cite{Denef:2000nb,Bellorin:2006xr}, this condition will lead to 
non-trivial constraints.

The vector field strengths are given by

\begin{equation}
F = -{\textstyle\frac{1}{\sqrt{2}}} \{d[|M|^{2}\mathcal{R}(dt+\omega) ] 
-{}^{\star}[ |M|^{2} d\mathcal{I}\wedge(dt+\omega)] \}\, ,
\end{equation}

\noindent
and the scalar fields $Z^{i}$ can be computed by taking the quotients

\begin{equation}
Z^{i}=(\mathcal{V}/X)^{i}/(\mathcal{V}/X)^{0}\, .
\end{equation}

The hyperscalars $q^{u}(x)$ are just the mapping whose existence we
assumed from the onset. 

These solutions can therefore be seen as deformations of those devoid of hypers,
originally found in Ref.~\cite{Behrndt:1997ny}.

As for the number of unbroken supersymmetries, the presence of non-trivial
hyperscalars breaks $1/2$ or $1/4$ of the supersymmetries of the related
solution without hypers, which may have all or $1/2$ of the original
supersymmetries. Therefore, we will have solutions with $1/2,1/4$ and $1/8$ of
the original supersymmetries. The Killing spinors take the form

\begin{equation}
\epsilon_{I}=X^{1/2}\epsilon_{I\, 0}\, ,
\hspace{1cm}
\partial_{\mu}\epsilon_{I\, 0}=0\, ,
\hspace{1cm}  
\epsilon_{I\, 0} +i\gamma_{0}
  \varepsilon_{IJ}\epsilon^{J}{}_{0}=0\, ,
\hspace{1cm}  
\Pi^{x}{}_{I}{}^{J}\ \epsilon_{J\, 0} =0\, ,
\end{equation}

\noindent
where the first constraint is imposed only if there are non-trivial vector
multiplets and each of the other three constraints is imposed for each
non-vanishing component of the $SU(2)$ connection. Each constraint breaks
$1/2$ of the supersymmetries independently, but the third constraint
$\Pi^{x}{}_{I}{}^{J}\ \epsilon_{J\, 0} =0$ is implied by the first two.
Finally, the meaning of these last three constraints is that they enforce the
embedding of the gauge connection into the gauge connection since they are in
different representations.

\item In the null case the hyperscalars can only depend on the null coordinate
  $u$ and the solutions take essentially the same form as in the case without
  hypermultiplets (See Ref.~\cite{Meessen:2006tu}).

\end{enumerate}
\par
The plan of this article is as follows: in Section~\ref{sec-thetheory} we will
discuss the theory we are dealing with and especially the supersymmetry
transformations. This is followed in Sec.~\ref{sec-setup} by a short
discussion of the Killing spinor identities and their implications.
Secs.~\ref{sec-timelike} and \ref{sec-null} then deal with the explicit
solutions in the two possible cases that, according to the KSIs can, occur.
Finally, Appendix~\ref{sec-QKG} is devoted to quaternionic K\"ahler
geometry\footnote{ Our conventions, including those for the special K\"ahler
  geometry, are those of Ref.~\cite{Meessen:2006tu}.  } and
Appendix~\ref{appsec:dualquat} spells out the details for the c-map alluded to 
in the introduction.

\section{Matter-coupled $N=2$ $d=4$ ungauged supergravity}
\label{sec-thetheory}

The theory we are working with is an extension of the one studied in
Ref.~\cite{Meessen:2006tu}, the extension consisting in the additional
coupling of $m$ hypermultiplets. We refer the reader to
\cite{Meessen:2006tu} for all conventions and notations except for
those involving the $m$ hypermultiplets which we explain next. These
are essentially those of Ref.~\cite{Andrianopoli:1996cm} with the
minor changes introduced in Ref.~\cite{Meessen:2006tu}. Each
hypermultiplet consists of 4 real scalars $q$ (\textit{hyperscalars})
and 2 Weyl spinors $\zeta$ called \textit{hyperinos}. The $4m$
hyperscalars are collectively denoted by $q^{u}\,
,\,\,\,u=1,\cdots,4m$ and the $2m$ hyperinos are collectively denoted
by $\zeta_{\alpha}\, ,\,\,\,\alpha=1,\cdots,2m$.  The $4m$
hyperscalars parametrize a quaternionic K\"ahler manifold (defined and
studied in Appendix~\ref{sec-QKG}) with metric $\mathsf{H}_{uv}(q)$.

The action of the bosonic fields of the theory is

\begin{equation}
\label{eq:action}
\begin{array}{rcl}
 S & = & {\displaystyle\int} d^{4}x \sqrt{|g|}
\left[R +2\mathcal{G}_{ij^{*}}\partial_{\mu}Z^{i}
\partial^{\mu}Z^{*\, j^{*}} +2\mathsf{H}_{uv}\partial_{\mu}q^{u} \partial^{\mu}q^{v}
 \right. \\
& & \\
& & \left. 
\hspace{2cm}
+2\Im{\rm m}\mathcal{N}_{\Lambda\Sigma} 
F^{\Lambda\, \mu\nu}F^{\Sigma}{}_{\mu\nu}
-2\Re{\rm e}\mathcal{N}_{\Lambda\Sigma} 
F^{\Lambda\, \mu\nu}{}^{\star}F^{\Sigma}{}_{\mu\nu}
\right]\, ,
\end{array}
\end{equation}

For vanishing fermions, the supersymmetry transformation rules of the
fermions are

\begin{eqnarray}
\delta_{\epsilon}\psi_{I\, \mu} & = & 
\mathfrak{D}_{\mu}\epsilon_{I}
+\varepsilon_{IJ}\ T^{+}{}_{\mu\nu}\gamma^{\nu}\ \epsilon^{J}\, ,
\label{eq:gravisusyrule}\\
& & \nonumber \\
\delta_{\epsilon}\lambda^{iI} & = & 
i\not\!\partial Z^{i}\epsilon^{I} \ +\ \varepsilon^{IJ}\not\!G^{i\, +}\ \epsilon_{J}\, .
\label{eq:gaugsusyrule} \\
& & \nonumber \\
\delta_{\epsilon}\zeta_{\alpha} & = &
-i\mathbb{C}_{\alpha\beta}\ \mathsf{U}^{\beta I}{}_{u}\ \varepsilon_{IJ}\ 
\not\!\partial q^{u}\ \epsilon^{J}\, , 
\label{eq:hypersusyrule}
\end{eqnarray}

\noindent
Here $\mathfrak{D}$ is the Lorentz and K\"ahler-covariant
derivative of Ref.~\cite{Meessen:2006tu} supplemented by (the pullback of) an
$SU(2)$ connection $\mathsf{A}_{I}{}^{J}$ described in
Appendix~\ref{sec-QKG}, acting on objects with $SU(2)$ indices
$I,J$ and, in particular, on $\epsilon_{I}$ as:

\begin{equation}
\mathfrak{D}_{\mu} \epsilon_{I} = 
(\nabla_{\mu} \ +\ {\textstyle\frac{i}{2}}\ \mathcal{Q}_{\mu})\ \epsilon_{I} 
\ +\ \mathsf{A}_{\mu\, I}{}^{J}\ \epsilon_{J}\, .
\end{equation}

\noindent
This is the only place in which the hyperscalars appear in the
supersymmetry transformation rules of the gravitinos and gauginos.
$\mathsf{U}^{\beta I}{}_{u}$ is a Quadbein, {\em i.e.\/} a quaternionic Vielbein, 
and $\mathbb{C}_{\alpha\beta}$ the $Sp(m)$-invariant metric,
both of which are described in Appendix~\ref{sec-QKG}.

The supersymmetry transformations of the bosons are the same as in the
previous case plus that of the hyperscalars:

\begin{eqnarray}
  \delta_{\epsilon} e^{a}{}_{\mu} & = & 
-{\textstyle\frac{i}{4}} (\bar{\psi}_{I\, \mu}\gamma^{a}\epsilon^{I}
+\bar{\psi}^{I}{}_{\mu}\gamma^{a}\epsilon_{I})\, ,
\label{eq:susytranse}\\
& & \nonumber \\ 
  \delta_{\epsilon} A^{\Lambda}{}_{\mu} & = & 
{\textstyle\frac{1}{4}}
(\mathcal{L}^{\Lambda\, *}
\varepsilon^{IJ}\bar{\psi}_{I\, \mu}\ \epsilon_{J}
\ +\
\mathcal{L}^{\Lambda}
\varepsilon_{IJ}\bar{\psi}^{I}{}_{\mu}\ \epsilon^{J}) \nonumber \\
& & \nonumber \\
& & 
+
{\textstyle\frac{i}{8}}(f^{\Lambda}{}_{i}\varepsilon_{IJ}
\bar{\lambda}^{iI}\gamma_{\mu}
\epsilon^{J}
+
f^{\Lambda *}{}_{i^{*}}\varepsilon^{IJ}
\bar{\lambda}^{i^{*}}{}_{I}\gamma_{\mu}\epsilon_{J})\, ,
\label{eq:susytransA}\\
& & \nonumber \\
  \delta_{\epsilon} Z^{i} & = & 
{\textstyle\frac{1}{4}} \bar{\lambda}^{iI}\epsilon_{I}\, ,
\label{eq:susytransZ}\\
& & \nonumber \\
  \delta_{\epsilon} q^{u} & = & \mathsf{U}_{\alpha I}{}^{u} 
(\bar{\zeta}^{\alpha}\epsilon^{I} 
+\mathbb{C}^{\alpha\beta}\epsilon^{IJ}\bar{\zeta}_{\beta}\epsilon_{J})\, .
\label{eq:susytransq}
\end{eqnarray}

\noindent
Observe that the fields of the hypermultiplet and the fields of the
gravity and vector multiplets do not mix in any of these
supersymmetry transformation rules. This means that the KSIs
\cite{Kallosh:1993wx,Bellorin:2005hy} associated to the gravitinos and
gauginos will have the same form as in Ref.~\cite{Meessen:2006tu} and in the
KSIs associated to the hyperinos only the hyperscalars equations of
motion will appear.

For convenience, we denote the bosonic equations of motion by

\begin{equation}
\mathcal{E}_{a}{}^{\mu}\equiv 
-\frac{1}{2\sqrt{|g|}}\frac{\delta S}{\delta e^{a}{}_{\mu}}\, ,
\hspace{.4cm}
\mathcal{E}_{i} \equiv -\frac{1}{2\sqrt{|g|}}
\frac{\delta S}{\delta Z^{i}}\, ,
\hspace{.4cm}
\mathcal{E}_{\Lambda}{}^{\mu}\equiv 
\frac{1}{8\sqrt{|g|}}\frac{\delta S}{\delta A^{\Lambda}{}_{\mu}}\, ,
\hspace{.4cm}
\mathcal{E}^{u}\equiv -\frac{1}{4\sqrt{|g|}}\mathsf{H}^{uv}\frac{\delta S}{\delta q^{v}}\, .
\end{equation}

\noindent
and the Bianchi identities for the vector field strengths by

\begin{equation}
\label{eq:BL}
\mathcal{B}^{\Lambda\, \mu} \equiv \nabla_{\nu}{}^{\star}F^{\Lambda\,
  \nu\mu}\, .  
\end{equation}

Then, using the action Eq.~(\ref{eq:action}), we find that all the
equations of motion of the bosonic fields of the gravity and vector
supermultiplets take the same form as if there were no
hypermultiplets, as in Ref.~\cite{Meessen:2006tu}, except for the Einstein
equation, which obviously is supplemented by the energy-momentum tensor of the
hyperscalars

\begin{equation}
\label{eq:EOMGrav}
\mathcal{E}_{\mu\nu}  \; =\;  \mathcal{E}_{\mu\nu}(q=0) 
\ +\ 2\mathsf{H}_{uv}\ [\partial_{\mu}q^{u}\partial_{\nu}q^{v} 
-{\textstyle\frac{1}{2}}g_{\mu\nu}\partial_{\rho}q^{u}\partial_{\rho}q^{v}]\, .
\end{equation}

\noindent
Furthermore, the equation of motion for the hyperscalars reads

\begin{equation}
\label{eq:EOMMierdini}
\mathcal{E}^{u}=\mathfrak{D}_{\mu}\partial^{\mu}q^{u}=
\nabla_{\mu}\partial^{\mu}q^{u} 
+\Gamma_{vw}{}^{u}\partial^{\mu}q^{v}\partial_{\mu}q^{w}\, ,  
\end{equation}
where $\Gamma_{vw}{}^{u}$ are the Christoffel symbols of the 2$^{nd}$ kind
for the metric $\mathsf{H}_{uv}$.
\par
The symmetries of this set of equations of motion are the isometries
of the K\"ahler manifold parametrized by the $\bar{n}-1$ complex
scalars $Z^{i}$s embedded in $Sp(2\bar{n},\mathbb{R})$ and those of
the quaternionic manifold parametrized by the $4m$ real scalars
$q^{u}$.


\section{Supersymmetric configurations: generalities}
\label{sec-setup}

As we mentioned in Section~\ref{sec-thetheory} the supersymmetry
transformation rules of the bosonic fields indicate that the KSIs
associated to the gravitinos and gauginos are going to have the same
form as in absence of hypermultiplets. This is indeed the case, and
the integrability conditions of the KSEs
$\delta_{\epsilon}\psi_{I\mu}=0$ and $\delta_{\epsilon}\lambda^{iI}=0$
confirm the results. Of course, now the Einstein equation includes an
additional term: the hyperscalars energy-momentum tensor. In the KSI
approach the origin of this term is clear. In the integrability
conditions it appears through the curvature of the $SU(2)$ connection
and Eq.~(\ref{eq:su2curvaturevsquaternionivielbeins}). The results
coincide for $\lambda=-1$.

There is one more set of KSIs associated to the hyperinos which take
the form

\begin{equation}
\label{eq:hyperKSI}
\mathcal{E}^{u}\ \mathsf{U}^{\alpha I}{}_{u}\ \epsilon_{I}\; =\; 0\,  ,  
\end{equation}

\noindent 
and which can be obtained from the integrability condition
$\not\!\!\mathfrak{D} \delta_{\epsilon}\zeta_{\alpha}=0$ using the
covariant constancy of the Quadbein, Eq. (\ref{eq:FullWax}).


The KSIs involving the equations of motion of the bosonic fields of the
gravity and vector multiplets take, of course, the same form as in absence of
hypermultiplets. Acting with $\bar{\epsilon}^{J}$ from the left on the new KSI
Eq.~(\ref{eq:hyperKSI}) we get

\begin{equation}
X \mathcal{E}^{u}\ \mathsf{U}^{\alpha I}{}_{u} \; =\; 0\, ,  
\end{equation}

\noindent
which implies, in the timelike $X\neq 0$ case, that all the supersymmetric 
configurations satisfy the hyperscalars equations of motion automatically:

\begin{equation}
\mathcal{E}^{u}=0\, .  
\end{equation}

In the null case, parametrizing the Killing spinors by
$\epsilon_{I}=\phi_{I}\epsilon$, we get just

\begin{equation}
\label{eq:ksinull6}
\mathcal{E}^{u}\ \mathsf{U}^{\alpha I}{}_{u}\ \phi_{I}\ \epsilon \; =\; 0\, .  
\end{equation}


As usual, there are two separate cases to be considered: the one in
which the vector bilinear $V^{\mu}\equiv i
\bar{\epsilon}^{I}\gamma^{\mu}\epsilon_{I}$, which is always going to
be Killing, is timelike (Section~\ref{sec-timelike}) and the one in
which it is null (Section~\ref{sec-null}).  The procedure we are going
to follow is almost identical to the one we followed in
Ref.~\cite{Meessen:2006tu}.


\section{The timelike case}
\label{sec-timelike}


As mentioned before, the presence of hypermultiplets only introduces
an $SU(2)$ connection in the covariant derivative
$\mathfrak{D}_{\mu}\epsilon_{I}$ in $\delta_{\epsilon}\psi_{I\mu}=0$
and has no effect on the KSE $\delta_{\epsilon}\lambda^{iI}=0$.
Following the same steps as in Ref.~\cite{Meessen:2006tu}, by
way of the gravitino
supersymmetry transformation rule Eq.~(\ref{eq:gravisusyrule}), we 
arrive at

\begin{eqnarray}
\mathfrak{D}_{\mu}X & = & -i T^{+}{}_{\mu\nu}V^{\nu}  \, ,
\label{eq:TV}\\
& & \nonumber \\
\mathfrak{D}_{\mu} V_{J}{}^{I}{}_{\nu} & = & 
i \delta^{I}{}_{J} (XT^{*-}{}_{\mu\nu} -X^{*}T^{+}{}_{\mu\nu})
-i(\epsilon^{IK}T^{*\, -}{}_{\mu\rho}\Phi_{KJ}{}^{\rho}{}_{\nu}
-\epsilon_{JK}T^{+}{}_{\mu\rho}\Phi^{IK}{}_{\nu}{}^{\rho})\, .
\end{eqnarray}

The $SU(2)$ connection does not occur in the first equation, simply because
$X=\frac{1}{2}\epsilon^{IJ}M_{IJ}$ is an $SU(2)$ scalar, but it does
occur in the second, although not in its trace. This means that
$V^{\mu}$ is, once again, a Killing vector and the 1-form $\hat{V}=V_{\mu}dx^{\mu}$
satisfies the equation

\begin{equation}
\label{eq:dV}
d\hat{V} = 4i (XT^{*\, -} -X^{*}T^{+})\, .
\end{equation}

\noindent
The remaining 3 independent 1-forms\footnote{
   $\sigma_{x\, J}{}^{I}\, ,\,\,\, (x=1,2,3)$ are the Pauli matrices
   satisfying Eq.~(\ref{eq:Paulimatrices}).
}

\begin{equation}
\hat{V}^{x} \;\equiv\;
{\textstyle\frac{1}{\sqrt{2}}} (\sigma_{x})_{I}{}^{J}\  V_{J}{}^{I}{}_{\mu}\ dx^{\mu}\, ,
\end{equation}

\noindent
however, are only $SU(2)$-covariantly exact

\begin{equation}
\label{eq:dVAV}
d\hat{V}^{x} \ +\ \varepsilon^{xyz}\ \mathsf{A}^{y}\wedge\hat{V}^{z}\; =\; 0\, .
\end{equation}

From $\delta_{\epsilon}\lambda^{iI}=0$ we get exactly the same
equations as in absence of hypermultiplets. In particular

\begin{eqnarray}
  \label{eq:VdZ}
V^{\mu}\partial_{\mu}Z^{i} & = & 0\, ,  \\
& & \nonumber \\
\label{eq:GV}
2i X^{*}\partial_{\mu}Z^{i}+4i G^{i\, +}{}_{\mu\nu}V^{\nu} & = & 0\, .  
\end{eqnarray}

Combine Eqs.~(\ref{eq:TV}) and (\ref{eq:GV}), we get

\begin{equation}
\label{eq:FV}
V^{\nu}F^{\Lambda\, +}{}_{\nu\mu}= 
\mathcal{L}^{*\, \Lambda}\mathfrak{D}_{\mu}X
+X^{*} f^{\Lambda}{}_{i}\partial_{\mu}Z^{i} =
 \mathcal{L}^{*\, \Lambda}\mathfrak{D}_{\mu}X
+X^{*}\mathfrak{D}_{\mu} \mathcal{L}^{\Lambda}\, ,
\end{equation}

\noindent
which, in the timelike case at hand, is enough to completely determine
through the identity

\begin{equation}
\label{eq:decomposition2}
C^{\Lambda\, +}{}_{\mu}\equiv V^{\nu}
F^{\Lambda\, +}{}_{\nu\mu}\,\,\, \Rightarrow\,\,\,
F^{\Lambda\, +}=V^{-2}[\hat{V}\wedge \hat{C}^{\Lambda\, +}
+ i\,{}^{\star}\!(\hat{V}\wedge \hat{C}^{\Lambda\, +})]\, .
\end{equation}

Observe that this equation does not involve the
hyperscalars in any explicit way, as was to be expected 
due to the absence of couplings
between the vector fields and the hyperscalars.

Let us now consider the new equation
$\delta_{\epsilon}\zeta_{\alpha}=0$. Acting on it from the left with
$\bar{\epsilon}^{K}$ and $\bar{\epsilon}^{K}\gamma_{\mu}$ we get,
respectively

\begin{eqnarray}
\label{eq:hyperkse1}
\mathsf{U}^{\alpha I}{}_{u}\ \varepsilon_{IJ}\ V^{J}{}_{K}{}^{\mu}\ \partial_{\mu} q^{u}
& = & 0\, ,\\
& & \nonumber \\
\label{eq:hyperkse2}
X^{*}\mathsf{U}^{\alpha K}{}_{u}\ \partial_{\mu}q^{u}
\, +\,\mathsf{U}^{\alpha I}{}_{u}\ \varepsilon_{IJ}\ 
\Phi^{KJ}{}_{\mu}{}^{\rho}\ \partial_{\rho}q^{u} & = & 0\, .
\end{eqnarray}

Using $\varepsilon_{IJ}\ V^{J}{}_{K} \ =\ \varepsilon_{KJ}\ V^{J}{}_{I}
\ +\ \varepsilon_{IK}\ V$ in the first equation we get

\begin{equation}
\label{eq:hyperkse3}
\mathsf{U}^{\alpha I}{}_{u}\ V^{J}{}_{I}{}^{\mu}\partial_{\mu} q^{u}
\ -\ \mathsf{U}^{\alpha J}{}_{u}\ V^{\mu}\partial_{\mu} q^{u} \; =\; 0\, .
\end{equation}

It is not difficult to see that the second equation can be derived
from this one using the Fierz identities that the bilinears satisfy in
the timelike case (see Ref.~\cite{Bellorin:2005zc}), whence the
only equations to be solved are (\ref{eq:hyperkse3}).


\subsection{The metric}
\label{sec-metric}

If we define the time coordinate $t$ by

\begin{equation}
V^{\mu}\partial_{\mu}\equiv\sqrt{2} \partial_{t}\, ,
\end{equation}

\noindent
then $V^{2}=4|X|^{2}$ implies that $\hat{V}$ must take the form

\begin{equation}
\hat{V}=2\sqrt{2}|X|^{2}(dt+\omega)\, ,  
\end{equation}

\noindent
where $\omega$ is a 1-form to be determined later.

Since the $\hat{V}^{x}$s are not exact, we cannot simply define coordinates
by putting $\hat{V}^{x}\equiv dx^{x}$. We can, however, still use them to
construct the metric: using

\begin{equation}
g_{\mu\nu}=2V^{-2}[V_{\mu}V_{\nu}-V_{J}{}^{I}{}_{\mu} V_{I}{}^{J}{}_{\nu}]\, ,
\end{equation}

\noindent
and the decomposition

\begin{equation}
\label{eq:vectordecomposition}
V_{J}{}^{I}{}_{\mu} \; =\; {\textstyle\frac{1}{2}}V_{\mu}\ \delta_{J}{}^{I} 
\; +\; {\textstyle\frac{1}{\sqrt{2}}}\ 
(\sigma_{x})_{J}{}^{I}\ V^{x}{}_{\mu}\, ,  
\end{equation}

\noindent
we find that the metric can be written in the form

\begin{equation}
ds^{2}= \frac{1}{4|X|^{2}}\hat{V}\otimes \hat{V} 
-\frac{1}{2|X|^{2}} \delta_{xy}\hat{V}^{x}\otimes \hat{V}^{y}\, .
\end{equation}

\noindent
The $\hat{V}^{x}$ are mutually orthogonal and also orthogonal to
$\hat{V}$, which means that they can be used as a Dreibein for a
3-dimensional Euclidean metric

\begin{equation}
\delta_{xy}\hat{V}^{x}\otimes \hat{V}^{y} \; \equiv\; 
\gamma_{\underline{m}\underline{n}}dx^{m}dx^{n}\, ,  
\end{equation}

\noindent
and the 4-dimensional metric takes the form

\begin{equation}
\label{eq:timelikemetric}
ds^{2} \; =\; 2|X|^{2} (dt+\omega)^{2} 
-\frac{1}{2|X|^{2}}\gamma_{\underline{m}\underline{n}}dx^{m}dx^{n}\, .  
\end{equation}

The presence of a non-trivial Dreibein and the corresponding 3D metric
$\gamma_{\underline{m}\underline{n}}$ is the main (and only)
novelty brought about by the hyperscalars!

In what follows we will use the Vierbein basis

\begin{equation}
e^{0}  =  \frac{1}{2|X|} \hat{V}\, ,
\hspace{1cm}
e^{x}  =  \frac{1}{\sqrt{2}|X|}\hat{V}^{x}\, ,
\end{equation}

\noindent
that is

\begin{equation}
\label{eq:Vierbeins}
(e^{a}{}_{\mu}) = 
\left(
  \begin{array}{cc}
\sqrt{2}|X| & \sqrt{2}|X| \omega_{\underline{m}} \\
& \\
0 & \frac{1}{\sqrt{2}|X|} V^{x}{}_{\underline{m}} \\
  \end{array}
\right)\, ,
\hspace{1cm}
(e^{\mu}{}_{a}) = 
\left(
  \begin{array}{cc}
\frac{1}{\sqrt{2}|X|} & -\sqrt{2}|X| \omega_{x} \\
& \\
0 & \sqrt{2}|X| V_{x}{}^{\underline{m}} \\
  \end{array}
\right)\, .
\end{equation}

\noindent
where $V_{x}{}^{\underline{m}}$ is the inverse Dreibein
$V_{x}{}^{\underline{m}} V^{y}{}_{\underline{m}}=\delta^{y}{}_{x}$ and
$ \omega_{x} = V_{x}{}^{\underline{m}}\omega_{\underline{m}}$.  We shall
also adopt the convention that all objects with flat or
curved 3-dimensional indices refer to the above Dreibein and the
corresponding metric.

Our choice of time coordinate Eq.~(\ref{eq:VdZ}) means that
the scalars $Z^{i}$ are time-independent, whence $\imath_{V}\mathcal{Q}\ =\ 0$. 
%
%
Contracting Eq.~(\ref{eq:TV}) with $V^{\mu}$ we get

\begin{equation}
V^{\mu}\mathfrak{D}_{\mu}\ X=0\, ,\,\,\, \Rightarrow\;\; V^{\mu}\partial_{\mu}X=0\, ,
\end{equation}

\noindent
so that also $X$ is time-independent.

We know the $\hat{V}^{x}$s to have no time components. If we choose
the gauge for the pullback of the $SU(2)$ connection
$\mathsf{A}^{x}{}_{t}=0$, then the $SU(2)$-covariant constancy of the
$\hat{V}^{x}$ (Eq.~(\ref{eq:dVAV})) states that the pullback of
$\mathsf{A}^{x}$, the $\hat{V}^{x}$s and, therefore, the
3-dimensional metric $\gamma_{\underline{m}\underline{n}}$ are also
time-independent.  Eq.~(\ref{eq:dVAV}) can then be interpreted as 
Cartan's first structure equation for a torsionless connection $\varpi$
in 3-dimensional  space

\begin{equation}
d\hat{V}^{x}-\varpi^{xy}\wedge\hat{V}^{y} =0\, ,
\end{equation}

\noindent
which means that the 3-dimensional spin connection 1-form
$\varpi_{x}{}^{y}$ is related to the pullback of the $SU(2)$ connection
$\mathsf{A}^{x}$ by

\begin{equation}
\label{eq:connectionemberdding}
\varpi_{\underline{m}}{}^{xy} = \varepsilon^{xyz}
\mathsf{A}^{z}{}_{u}\ \partial_{\underline{m}}q^{u}\, ,
\end{equation}

\noindent
implying the embedding of the internal group $SU(2)$ into the Lorentz
group of the 3-dimensional space as discussed in the introduction.

The $\mathfrak{su}(2)$ curvature will also be time-independent
and Eq.~(\ref{eq:su2curvaturevsquaternionivielbeins}) implies that the
pullback of the Quadbein is also time-independent and its
time component vanishes:

\begin{equation}
\label{eq:VU}
\mathsf{U}^{\alpha I}{}_{u}\  V^{\mu} \partial_{\mu}q^{u} =0\, .
\end{equation}

Let us then consider the 1-form $\omega$: following the same steps as
in Ref.~\cite{Meessen:2006tu}, we arrive at 

\begin{equation}
\label{eq:do23-d}
(d\omega)_{xy} =-\frac{i}{2|X|^{4}}\varepsilon_{xyz}
(X^{*}\mathfrak{D}^{z}X-X\mathfrak{D}^{z}X^{*})\, .  
\end{equation}

\noindent
This equation has the same form as in the case without hypermultiplets,
but now the Dreibein is non-trivial and, in curved indices, it takes the
form

\begin{equation}
(d\omega)_{\underline{m}\underline{n}} =-\frac{i}{2|X|^{4}\sqrt{|\gamma|}}
\varepsilon_{\underline{m}\underline{n}\underline{p}}
(X^{*}\mathfrak{D}^{\underline{p}}X-X\mathfrak{D}^{\underline{p}}X^{*})\, .  
\end{equation}

\noindent
Introducing the real symplectic sections $\mathcal{I}$ and
$\mathcal{R}$

\begin{equation}
\label{eq:realsections}
\mathcal{R}\equiv \Re{\rm e}(\mathcal{V}/X)\, ,
\hspace{1.5cm}
\mathcal{I}\equiv \Im{\rm m}(\mathcal{V}/X)\, ,
\end{equation}

\noindent
where $\mathcal{V}$ is the symplectic section

\begin{equation}
\label{eq:SGDefFund}
\mathcal{V} = 
\left( \!
\begin{array}{c}
\mathcal{L}^{\Lambda}\\
\mathcal{M}_{\Sigma}\\
\end{array}
\!\right)\, ,
\hspace{1cm}
\langle \mathcal{V}\mid\mathcal{V}^{*}\rangle 
 \equiv  
\mathcal{L}^{*\, \Lambda}\mathcal{M}_{\Lambda} 
-\mathcal{L}^{\Lambda}\mathcal{M}^{*}_{\Lambda}
= -i\, ,
\end{equation}

\noindent
we can rewrite the equation for $\omega$ to the alternative form

\begin{equation}
\label{eq:oidi}
(d\omega)_{xy} =2 \epsilon_{xyz}
\langle\,\mathcal{I}\mid \partial^{z}\mathcal{I}\, \rangle\, ,  
\end{equation}

\noindent
whose integrability condition is

\begin{equation}
\langle\,\mathcal{I}\mid \nabla_{\underline{m}}\partial^{\underline{m}}
\mathcal{I}\, \rangle =0\, ,    
\end{equation}

\noindent
and will be satisfied by harmonic functions on the 3-dimensional space,
{\em i.e.} by those real symplectic sections satisfying
$\nabla_{\underline{m}}\partial^{\underline{m}} \mathcal{I}=0$. In
general the harmonic functions will have singularities leading to
non-trivial constraints like those studied in
Refs.~\cite{Denef:2000nb,Bellorin:2006xr}.


\subsection{Solving the Killing spinor equations}
\label{sec-solvingKSEtimelike}

We are now going to see that it is always possible to solve the KSEs
for field configurations with metric of the form
(\ref{eq:timelikemetric}) where the 1-form $\omega$ satisfies
Eq.~(\ref{eq:do23-d}) and the 3-dimensional metric has spin connection
related to the $SU(2)$ connection by
Eq.~(\ref{eq:connectionemberdding}), vector fields of the form
(\ref{eq:FV}) and (\ref{eq:decomposition2}), time-independent scalars
$Z^{i}$ and, most importantly, hyperscalars satisfying

\begin{equation}
\label{eq:hyperkse4}
\mathsf{U}^{\alpha J}{}_{x}\ (\sigma_{x})_{J}{}^{I} \; =\; 0\, ,
\hspace{1cm}
\mathsf{U}^{\alpha J}{}_{x}\ \equiv\ V_{x}{}^{\underline{m}}\partial_{\underline{m}}q^{u}\
\mathsf{U}^{\alpha J}{}_{u}\, ,
\end{equation}

\noindent
which results from Eqs.~(\ref{eq:hyperkse3}), (\ref{eq:VU}) and
(\ref{eq:vectordecomposition}).

Let us consider first the $\delta_{\epsilon}\zeta_{\alpha}=0$
equation. Using the Vierbein Eq.~(\ref{eq:Vierbeins}) and multiplying
by $\gamma^{0}$ it can be rewritten in the form

\begin{equation}
  \mathsf{U}_{\alpha I\, x}\ \gamma^{0x}\ \epsilon^{I}\; =\; 0\, ,
\end{equation}

\noindent
which can be solved using Eq.~(\ref{eq:hyperkse4}) if the spinors
satisfy a constraint

\begin{equation}\label{eq:Pix}
\Pi^{x}{}_{I}{}^{J}\ \epsilon_{J}\; =\; 0 \hspace{.5cm} ,\hspace{.5cm}
\Pi^{x}{}_{I}{}^{J} \;\equiv\; 
{\textstyle\frac{1}{2}}
[\ \delta_{I}{}^{J}\ -\ \gamma^{0(x)}\ (\sigma_{(x)})_{I}{}^{J}\ ]\hspace{1cm}\mbox{(no sum over $x$)},
\end{equation}

\noindent
for each non-vanishing $\mathsf{U}_{\alpha I\, x}$.  These three operators are
projectors, {\em i.e.\/} they satisfy $(\Pi^{x})^{2}=\Pi^{x}$, and commute
with each other.  From $(\sigma_{(x)})_{I}{}^{K}\ \Pi^{(x)}{}_{K}{}^{J}\ 
\epsilon_{J}\ =\ 0$ we find

\begin{equation}
(\sigma_{(x)})_{I}{}^{J}\epsilon_{J} \; =\; \gamma^{0(x)} \epsilon_{I}\, ,
\end{equation}

\noindent
which solves $\delta_{\epsilon}\zeta_{\alpha}=0$ together with
Eq.~(\ref{eq:hyperkse4}) and tells us that the embedding of the
$SU(2)$ connection in the Lorentz group requires the action of the
generators of $\mathfrak{su}(2)$ to be identical to the action of the three Lorentz
generators $\frac{1}{2}\gamma^{0x}$ on the spinors. When we impose
these constraints on the spinors, each of the first two reduces by a
factor of $1/2$ the number of independent spinors, but the third
condition is implied by the first two and does not reduce any further
the number of independent spinors.

Observe that 

\begin{equation}
\label{eq:Pixcc}
\Pi^{x\, I}{}_{J}\; \equiv\; (\Pi^{x}{}_{I}{}^{J})^{*} \; =\; 
-\varepsilon^{IK}\ \Pi^{x}{}_{K}{}^{L}\ \varepsilon_{LJ}\, .
\end{equation}

Let us now consider the equation $\delta_{\epsilon}\lambda^{iI}=0$. It
takes little to no time to realize that it reduces to the same form as
in absence of hypermultiplets

\begin{equation}
\delta_{\epsilon}\lambda^{iI}\; =\; i\not\! \partial Z^{i}\ 
(\epsilon^{I}+i\gamma_{0}e^{-i\alpha} \varepsilon^{IJ}\epsilon_{J}\ )\; =0\; \, ,  
\end{equation}

\noindent
the only difference being in the implicit presence of the non-trivial
Dreibein in $\not\!\! \partial Z^{i}$. Therefore, as before, this equation is solved by
imposing the constraint

\begin{equation}
\epsilon^{I} \ +\ i\gamma_{0}\ e^{-i\alpha} \varepsilon^{IJ}\ \epsilon_{J}\; =\; 0\, ,
\end{equation}

\noindent
which can be seen to commute with the projections $\Pi^{x}$ since,  by virtue
of Eq.~(\ref{eq:Pixcc}),

\begin{equation}
\Pi^{x\, K}{}_{I}\ 
(\epsilon^{I}+i\gamma_{0}e^{-i\alpha} \varepsilon^{IJ}\epsilon_{J}) 
\; =\; (\Pi^{x\, K}{}_{I}\epsilon^{I})+i\gamma_{0}e^{-i\alpha} \varepsilon^{KJ}
(\Pi^{x}{}_{J}{}^{L}\epsilon_{L})\, .
\end{equation}

\noindent
Let us finally consider the equation
$\delta_{\epsilon}\lambda^{iI}=0$: in the $SU(2)$ gauge
$\mathsf{A}^{x}{}_{t}=0$ the $0$th component of the equation is
automatically solved by time-independent Killing spinors using the
above constraint. Again, the equation takes the same form as without
hypermultiplets but with a non-trivial Dreibein. In the same
gauge, the spatial (flat) components of the
$\delta_{\epsilon}\lambda^{iI}=0$ equation can be written, upon use of
the above constraint and the relation
Eq.~(\ref{eq:connectionemberdding}) between the $SU(2)$ and spatial
spin connection, in the form

\begin{equation}
  X^{1/2}\partial_{y}(X^{-1/2}\epsilon_{I}) 
  \ +\ {\textstyle\frac{i}{2}}\mathsf{A}^{x}{}_{y}\ 
  [(\sigma_{x})_{I}{}^{J}\epsilon_{J}-\gamma^{0x} \epsilon_{I}] \; =\; 0
  \hspace{.5cm},\hspace{.5cm}
  \mathsf{A}^{x}{}_{y}\ =\
  \mathsf{A}^{x}{}_{u} \partial_{\underline{m}}q^{u} \ V_{y}{}^{\underline{m}}\, ,
\end{equation}

\noindent
which is solved by 

\begin{equation}
\epsilon_{I}=X^{1/2}\epsilon_{I\, 0}\, ,
\hspace{1cm}
\partial_{\mu}\epsilon_{I\, 0}=0\, ,
\hspace{1cm}  
\epsilon_{I\, 0} +i\gamma_{0}
  \varepsilon_{IJ}\epsilon^{J}{}_{0}=0\, ,
\hspace{1cm}  
\Pi^{x}{}_{I}{}^{J}\ \epsilon_{J\, 0} =0\, ,
\end{equation}

\noindent
where the constraints Eq.~(\ref{eq:Pix}) are imposed for each
non-vanishing component of the $SU(2)$ connection.


\subsection{Equations of motion}
\label{sec-eoms}

According to the KSIs, all the equations of motion of the
supersymmetric solutions will be satisfied if the Maxwell equations
and Bianchi identities of the vector fields are satisfied. Before
studying these equations it is important to notice that supersymmetry
requires Eqs.~(\ref{eq:hyperkse4}) to be satisfied. We will assume
here that this has been done and we will study in the next section
possible solutions to these equations.

Using Eqs.~(\ref{eq:FV}) and (\ref{eq:decomposition2}) we can write
the symplectic vector of 2-forms in the form

\begin{equation}
F =\frac{1}{2|X|^{2}}
\{
\hat{V}\wedge d[|X|^{2}\mathcal{R}]
-{}^{\star}[\hat{V}\wedge \Im{\rm m}(\mathcal{V}^{*}\mathfrak{D}X 
+X^{*}\mathfrak{D}\mathcal{V})]
\} \, , 
\end{equation}

\noindent
which can be rewritten in the form

\begin{equation}
\label{eq:esas}
F = -{\textstyle\frac{1}{2}} \{d[\mathcal{R} \hat{V}] 
+{}^{\star}[\hat{V}\wedge  d\mathcal{I}] \}\, .
\end{equation}

The Maxwell equations and Bianchi identities $dF=0$ are, therefore,
satisfied if 

\begin{equation}
d{}^{\star}[\hat{V}\wedge  d\mathcal{I}] =0\, ,\,\,\, \Rightarrow
\nabla_{\underline{m}}\partial^{\underline{m}}\mathcal{I}=0\, ,
\end{equation}

\noindent
i.e.~if the $2\bar{n}$ components of $\mathcal{I}$ are as many real
harmonic functions in the 3-dimensional space with metric
$\gamma_{\underline{m}\underline{n}}$.

Summarizing, the timelike supersymmetric solutions are determined by a
choice of Dreibein and hyperscalars such that
Eq.~(\ref{eq:hyperkse4}) is satisfied and a choice of $2\bar{n}$ real
harmonic functions in the 3-dimensional metric space determined by our
choice of Dreibein $\mathcal{I}$. This choice determines the 1-form
$\omega$. The full $\mathcal{V}/X$ is determined in terms of
$\mathcal{I}$ by solving the stabilization equations and with
$\mathcal{V}/X$ one constructs the remaining elements of the solution
as explained in Ref.~\cite{Meessen:2006tu}.


\subsection{The cosmic string scrutinized}
\label{sec:CosmString}

It is always convenient to have an example that shows that we are not dealing
with an empty set of solutions. As mentioned in the introduction we can find
relatively simple non-trivial examples using the c-map on known supersymmetric
solutions with only fields in the vector multiplets excited. A convenient
solution is the cosmic string for the case $n=1$ with scalar manifold
$Sl(2,\mathbb{R})/U(1)$ and prepotential $\mathcal{F} =
-\textstyle{\frac{i}{4}} \mathcal{X}^{0}\mathcal{X}^{1}$.  Parametrizing the
scalars as $\mathcal{X}^{0}=1$ and $\mathcal{X}^{1}= -i\tau$, we find from
the formulae in appendix (\ref{appsec:dualquat}) that the only non-trivial
fields of the c-dual solution are the spacetime metric
\begin{equation}
  \label{eq:ScrutMet}
  ds^{2} \; =\; 2du\ dv \ -\ 2\ \mathrm{Im}(\tau)\ dz dz^{*} \; ,
\end{equation}
with $\tau = \tau (z)$, and the pull-back of the Quadbein is given
by
\begin{equation}
  \label{eq:ScrutQuad}
  \slashed{\mathsf{U}}^{\alpha I} \; =\; 
   \left[ 2\mathrm{Im}(\tau )\right]^{-3/2}\ \left(
     \begin{array}{lcl}
       0 & 0 \\ 0 & 0 \\ 
       \partial_{z}\tau\ \gamma^{z} & 0 \\
       0 & \partial_{z^{*}}\tau^{*}\ \gamma^{z^{*}}
     \end{array}
   \right) \; .
\end{equation}

From this form, then, it should be clear that the hyperscalar equation (\ref{eq:hypersusyrule}) is
satisfied by
\begin{equation}
  \label{eq:CosmString1}
  \gamma^{z}\epsilon^{2} \; =\; \gamma^{z^{*}}\epsilon^{1} \; =\; 0
     \hspace{.3cm}\longrightarrow\hspace{.3cm}
  \gamma^{z}\epsilon_{1} \; =\; \gamma^{z^{*}}\epsilon_{2} \; =\; 0\; ,
\end{equation}
so that we have to face the fact that this solution can be at most
1/2-BPS.
\par
Since we are dealing with a situation without vector multiplets and 
with a vanishing graviphoton, the gravitino variation (\ref{eq:gravisusyrule})
reduces to
\begin{equation}
  \label{eq:CosmString2}
  0\; =\; \nabla\ \epsilon_{I} \; +\; \mathsf{A}_{I}{}^{J}\epsilon_{J} \; .
\end{equation}
For the c-mapped cosmic string, we have from Eq. (\ref{eq:CMapSP1Con}),
that $\mathsf{A}_{I}{}^{J} = \textstyle{\frac{i}{2}}\ \mathcal{Q}\ \sigma_{3\ 
  I}{}^{J}$.  Also, for the metric at hand, the 4-d spin connection is readily
calculated to be $\textstyle{\frac{1}{2}}\omega_{ab}\gamma^{ab} =
i\mathcal{Q}\ \gamma^{zz^{*}}$ (See {\em e.g.\/} \cite{Bellorin:2005zc}).
\par
Due to the constraint (\ref{eq:CosmString1}), however, one can see that
$\gamma^{zz^{*}}\epsilon_{I} \; =\; \sigma_{3\ I}{}^{J}\ \epsilon_{J}$, which, when mixed
with the rest of the ingredients, leads to, dropping the $I$-indices,
\begin{equation}
  \mbox{Eq. }(\ref{eq:CosmString2}) \ =\ d\epsilon \ -\ \textstyle{\frac{1}{4}}\omega_{ab}\gamma^{ab}\epsilon \ +\ \textstyle{\frac{i}{2}}\mathcal{Q}\sigma_{3}\epsilon
 \ =\ d\epsilon \; ,
\end{equation}
so that the c-mapped cosmic string is a 1/2-BPS solution with, as was to be expected,
a constant Killing spinor.

\section{The null case}
\label{sec-null}

In the null case\footnote{
  The details concerning the normalization of the spinors and the 
  construction of the bilinears in this case are explained in the Appendix of
  Ref.~\cite{Bellorin:2005zc}, which you are strongly urged to consult at this point.} 
the two spinors $\epsilon_{I}$ are
proportional: $\epsilon_{I}=\phi_{I}\epsilon$. The complex
functions $\phi_{I}$, normalized such that $\phi^{I}\phi_{I}=1$
and satisfying $\phi_{I}^{*} =\phi^{I}$,
carry a -1 $U(1)$ charge {\em w.r.t.\/} the imaginary connection

\begin{equation}
  \zeta\; \equiv\;  \phi^{I}\ \mathfrak{D}\ \phi_{I}
  \hspace{.4cm}\rightarrow\hspace{.4cm} \zeta^{*} \; =\; -\zeta \, ,
\end{equation}

\noindent
opposite to that of the spinor $\epsilon$, whence $\epsilon_{I}$ is
neutral. On the other hand, the $\phi_{I}$s are neutral with respect
to the K\"ahler connection, and the K\"ahler weight of the spinor
$\epsilon$ is the same as that of the spinor $\epsilon_{I}$,
i.e.~$1/2$. The $SU(2)$-action is the one implied by the $I$-index
structure.

The substitution of the null-case spinor condition into the 
KSEs (\ref{eq:gravisusyrule}--\ref{eq:hypersusyrule}) 
immediately yields

\begin{eqnarray}
\mathfrak{D}_{\mu} \phi_{I} \epsilon+\phi_{I} \mathfrak{D}_{\mu}\epsilon 
+\varepsilon_{IJ}\phi^{J} T^{+}{}_{\mu\nu}
\gamma^{\nu} \epsilon^{*} & = & 0\, , 
\label{eq:gravitinodegenerate} \\
& & \nonumber \\
\phi^{I}\!\not\!\partial Z^{i} \epsilon^{*} +\varepsilon^{IJ}\phi_{J}
\not\! G^{i\, +} \epsilon & = & 0 \, , 
\label{eq:dilatinodegenerate}\\
& & \nonumber \\
\mathbb{C}_{\alpha\beta}\mathsf{U}^{\beta I}{}_{u}\varepsilon_{IJ}
\not\!\partial q^{u}\phi^{J}\epsilon^*&=&0.
\label{eq:hyperinodegenerate}
\end{eqnarray}

Contracting Eq.~(\ref{eq:gravitinodegenerate}) with $\phi^{I}$ results in 

\begin{equation}
\label{eq:Depsilon}
\mathfrak{D}_{\mu}\epsilon \; =\;  -\phi^{I}\ \mathfrak{D}_{\mu} \phi_{I}\ \epsilon
\;\;\longleftarrow\;\; 
\tilde{\mathfrak{D}}_{\mu}\epsilon \; \equiv\; 
(\mathfrak{D}_{\mu} \ +\ \zeta_{\mu})\epsilon \; =\; 0\, ,  
\end{equation}

\noindent
which is the only differential equation for $\epsilon$.
Substituting
Eq.~(\ref{eq:Depsilon}) into Eq.~(\ref{eq:gravitinodegenerate}) as to
eliminate the $\mathfrak{D}_{\mu}\epsilon$ term, we obtain

\begin{equation}
\label{eq:gravitinodegII}
\left(\tilde{\mathfrak{D}}_{\mu}\phi_{I}\right)\ 
\epsilon  \ +\ \varepsilon_{IJ}\phi^{J}\  T^{+}{}_{\mu\nu}\gamma^{\nu}\ 
\epsilon^{*}\; =\; 0\, ,
\hspace{2cm}
\tilde{\mathfrak{D}}_{\mu}\phi_{I} \equiv (\mathfrak{D}_{\mu} -\zeta_{\mu})\phi_{I}\, ,
\end{equation}

\noindent
which is a differential equation for $\phi_{I}$ and, at the same time,
an algebraic constraint for $\epsilon$.
Two further algebraic constraints can be found by acting with $\phi^{I}$ 
on Eq.~(\ref{eq:dilatinodegenerate}):

\begin{eqnarray}
\slashed{\partial} Z^{i}\, \epsilon^{*} \; = \;  
\slashed{G}^{i\, +}\ \epsilon \;  =\; 0\, . \label{eq:vinF-}
\end{eqnarray}

Finally, we add to the set-up an auxiliary spinor $\eta$, with the
same chirality as $\epsilon$ but with all $U(1)$ charges reversed,
and impose the normalization condition

\begin{equation}
\bar{\epsilon}\eta={\textstyle\frac{1}{2}}\, .  
\end{equation}

\noindent
This normalization condition will be preserved if and only if $\eta$
satisfies the differential equation

\begin{equation}
\label{eq:Deta}
\tilde{\mathfrak{D}}_{\mu}\eta \ +\ \mathfrak{a}_{\mu}\ \epsilon \; =\; 0\, ,  
\end{equation}

\noindent
for some $\mathfrak{a}$ with $U(1)$ charges $-2$ times those of $\epsilon$, i.e.

\begin{equation}
\tilde{\mathfrak{D}}_{\mu}\ \mathfrak{a}_{\nu} \, =\,
(\nabla_{\mu} \ -\ 2\zeta_{\mu} \ -\ i\mathcal{Q}_{\mu})\ \mathfrak{a}_{\nu}\, .
\end{equation}

\noindent
$\mathfrak{a}$ is to be determined by the requirement that the integrability conditions
of the above differential equation be compatible with those for $\epsilon$.


\subsection{Killing equations for the vector bilinears and first consequences}
We are now ready to derive equations involving the bilinears, in
particular the vector bilinears which we construct with $\epsilon$ and
the auxiliary spinor $\eta$ introduced above. First we deal with the
equations that do not involve derivative of the spinors. Acting with
$\bar{\epsilon}$ on Eq.~(\ref{eq:gravitinodegII}) and with
$\bar{\epsilon}\gamma^{\mu}$ on Eq.~(\ref{eq:vinF-}) we find

\begin{equation}
 T^{+}{}_{\mu\nu}\ l^{\nu} \; =\; 
 G^{i\, +}{}_{\mu\nu}\ l^{\nu} \; =\; 0 \hspace{.4cm}\longrightarrow\hspace{.4cm}
 F^{\Lambda\, +} {}_{\mu\nu}\ l^{\nu} \; =\; 0\, ,
\end{equation}

\noindent which implies
\begin{equation}
\label{eq:FL}
 F^{\Lambda\, +}\; =\; {\textstyle\frac{1}{2}}\ \varphi^{\Lambda}\ 
 \hat{l}\wedge \hat{m}^{*}\, ,   
\end{equation}

\noindent
for some complex functions $\varphi^{\Lambda}$. 
Acting with $\bar{\eta}$ on Eq.~(\ref{eq:gravitinodegII}) we get

\begin{equation}
\label{eq:gravitinodegIII}
\tilde{\mathfrak{D}}_{\mu}\phi_{I}
+i\sqrt{2}\varepsilon_{IJ}\phi^{J} T^{+}{}_{\mu\nu}m^{\nu}= 0\, ,
\end{equation}

\noindent
and substituting Eq.~(\ref{eq:FL}) into  it, we arrive at

\begin{equation}
\label{eq:gravitinodegIV}
\tilde{\mathfrak{D}}_{\mu}\ \phi_{I}
\ -\ {\textstyle\frac{i}{\sqrt{2}}} \ \varepsilon_{IJ}\phi^{J}\ \mathcal{T}_{\Lambda} 
\varphi^{\Lambda}\ l_{\mu}\; =\; 0\, .
\end{equation}

Finally, acting with $\bar{\epsilon}$ and $\bar{\eta}$ on
Eq.~(\ref{eq:vinF-}) we get

\begin{equation}
  \label{eq:dZ}
  l^{\mu}\ \partial_{\mu}Z^{i} \; =\; 
  m^{\mu}\ \partial_{\mu}Z^{i} \; =\; 0 \hspace{.3cm}\longrightarrow\hspace{.3cm}
  dZ^{i} \; =\; A^{i}\ \hat{l}\; +\; B^{i}\ \hat{m}\, ,
\end{equation}

\noindent
for some functions $A^{i}$ and $B^{i}$.

The relevant differential equations specifying the possible spacetime
dependencies for the tetrad follow from
Eqs.~(\ref{eq:Depsilon}) and (\ref{eq:Deta}). {\em I.e.\/}

\begin{eqnarray}
\label{eq:dtetrad1}
  \nabla_{\mu}\ l_{\nu} & = & 0\, ,\\
& & \nonumber \\
\label{eq:dtetrad2}
  \tilde{\mathfrak{D}}_{\mu}\ n_{\nu} &\equiv & \nabla_{\mu}\ n_{\nu} \; =\;
-\mathfrak{a}^{*}_{\mu}\ m_{\nu} \; -\; \mathfrak{a}_{\mu}\ m^{*}_{\nu}\, ,\\
& & \nonumber \\
\label{eq:dtetrad3}
  \tilde{\mathfrak{D}}_{\mu} m_{\nu} & \equiv & 
(\nabla_{\mu} \ -\ 2\zeta_{\mu} \ -\ i\mathcal{Q}_{\mu})\ m_{\nu} 
 \; =\; -\mathfrak{a}_{\mu}\ l_{\nu}\, .
\end{eqnarray}


\subsection{Equations of motion and integrability constraints}

As was discussed in Sec.~(\ref{sec-setup}), the KSIs in the case at
hand don't vary a great deal, with respect to the ones derived in 
\cite{Meessen:2006tu}, and so we can be brief: the only equations
of motion that are automatically satisfied are the ones for the 
graviphoton and the ones for the scalars from the vector multiplets.
As one can see from Eq. (\ref{eq:ksinull6}), the same thing cannot be
said about the equation of motion for the hyperscalar, but as we shall
see in a few pages, it is anyhow identically satisfied.
The, at the moment, relevant KSI is

\begin{equation}
  \label{eq:ksinull1}
  \left(\mathcal{E}_{\mu\nu} 
   -{\textstyle\frac{1}{2}}g_{\mu\nu}\, 
    \mathcal{E}_{\sigma}{}^{\sigma}
   \right)\ l^{\nu}  
   \; =\;
   \left( \mathcal{E}_{\mu\nu} 
      -{\textstyle\frac{1}{2}}g_{\mu\nu}\, \mathcal{E}_{\sigma}{}^{\sigma}
   \right)\ m^{\nu}
  \; =\;  0\, ,
\end{equation}

\noindent where the relation of the equation of motion with and without
hypermultiplets is given in Eq. (\ref{eq:EOMGrav}).

Substituting the expressions (\ref{eq:dZ}) and (\ref{eq:FL}) into the above KSIs 
we find the two conditions

\begin{eqnarray}
\label{eq:inte1}
0 & =& \left[\ 
           R_{\mu\nu} \ +\
           2\mathsf{H}_{uv}\ \partial_{\mu}q^{u}\ \partial_{\nu}q^{v}
       \right]\ l^{\nu} \, ,\\
& & \nonumber \\
\label{eq:inte2}
0 & =& \left[\
         R_{\mu\nu} \ +\
         2\mathsf{H}_{uv}\ \partial_{\mu}q^{u}\ \partial_{\nu}q^{v}
       \right]\ m^{\nu} 
       \ -\ \mathcal{G}_{ij^{*}}\ 
          \left(
             A^{i}l_{\mu}\ +\ B^{i}m_{\mu}
          \right)\ B^{*\, j^{*}} \, .
\end{eqnarray}

Comparable equations can be found from the integrability conditions of
Eq. (\ref{eq:Depsilon}), {\em i.e.\/}

\begin{eqnarray}
0 & =& \left[\  R_{\mu\nu} \ +\ 2 (d\zeta)_{\mu\nu}\right]\ l^{\nu} \, ,\\
& & \nonumber \\
0 & =& \left[\ R_{\mu\nu} \ +\ 2 (d\zeta)_{\mu\nu}\right]\  m^{*\, \nu}
  \ -\ \mathcal{G}_{ij^{*}}B^{i}\ (A^{*\, j^{*}}\ l_{\mu}
  \ +\ B^{*\, j^{*}}\ m^{*}_{\mu}) \label{eq:inte2b}\, ,
\end{eqnarray}

\noindent
and those of Eq.~(\ref{eq:Deta})

\begin{eqnarray}
0 & =& \left[\ 
          R_{\mu\nu} \ -\ 2 (d\zeta)_{\mu\nu} 
       \right]\ m^{\nu} 
  \ -\ \mathcal{G}_{ij^{*}}\ (A^{i}l_{\mu}+B^{i}m_{\mu})\ B^{*\, j^{*}}
  \ +\ 2(\tilde{\mathfrak{D}}\mathfrak{a})_{\mu\nu}l^{\nu} \, ,\\
& & \nonumber \\
0 & =& \left[\ 
         R_{\mu\nu} -2 (d\zeta)_{\mu\nu}
       \right]\ n^{\nu} 
  \ +\ 2(\tilde{\mathfrak{D}}\mathfrak{a})_{\mu\nu}\ m^{*\, \nu} \label{eq:ra}\, .
\end{eqnarray}

In the derivation of these last identities use has been made of the formulae

\begin{equation}
(d\mathcal{Q})_{\mu\nu}\ m^{* \nu} 
\; =\; i\mathcal{G}_{ij^{*}}\ B^{i}\ B^{*\, j^{*}}\ m^{*}_{\mu}\, ,
\hspace{1cm}  
(d\mathcal{Q})_{\mu\nu}\ l^{\nu}\; =\; (d\mathcal{Q})_{\mu\nu}\; n^{\nu}\; =\; 0\, ,
\end{equation}

\noindent
which follow from the definition of the K\"ahler connection and from
Eq.~(\ref{eq:dZ}).

Comparing these three sets of equations, we find that they are
compatible if

\begin{eqnarray}
  (d\zeta)_{\mu\nu}\ l^{\nu} & =& \mathsf{H}_{uv}\ \partial_{\mu}q^{u}\;
           l^{\nu}\partial_{\nu}q^{v} \; ,\label{eq:dzetal}\\
  (d\zeta)_{\mu\nu}\ m^{*\nu} & =& \mathsf{H}_{uv}\ \partial_{\mu}q^{u}\; 
           m^{*\nu}\partial_{\nu}q^{v} \label{eq:dzetamstar}\, ,
\end{eqnarray}
and
\begin{equation}
(\tilde{\mathfrak{D}}\ \mathfrak{a})_{\mu\nu}\; l^{\nu}\; =\; 0\; .
\end{equation}

\noindent Please observe that, due to the positive definiteness of $\mathsf{H}$,
Eq. (\ref{eq:dzetal}) implies $l^{\nu}\partial_{\nu}q^{v} =0$, but that
Eq. (\ref{eq:dzetamstar}) need not imply $m^{*\nu}\partial_{\nu}q^{v}=0$.


\subsection{A coordinate system, some more consistency and an anti-climax}

In order to advance in our quest, it is useful to introduce a coordinate
representation for the tetrad and hence also for the metric. 
Since $\hat{l}$ is a covariantly constant vector, we can introduce coordinates $u$
and $v$ through $l^{\mu}\partial_{\mu} = \partial_{v}$ and $l_{\mu}dx^{\mu} = du$.
We can also define a complex coordinates $z$ and $z^{*}$ by 

\begin{equation}
\label{eq:z}
\hat{m} \; =\;  e^{U}\ dz \hspace{.3cm},\hspace{.3cm}
\hat{m}^{*} \; =\; e^{U}\ dz^{*} \, ,  
\end{equation}

\noindent
where $U$ may depend on $z$, $z^{*}$ and $u$, but not $v$. Eq.~(\ref{eq:dZ})
then implies that the scalars $Z^{i}$ are just functions of $z$ and $u$:

\begin{equation}\label{eq:ziuz}
Z^{i}=Z^{i}(z,u)\, ,
\end{equation}

\noindent
wherefore the functions $A^{i}$ and $B^{i}$ defined in
Eq.~(\ref{eq:dZ}) are

\begin{equation}
\label{eq:functionsAB}
A^{i} \; =\; \partial_{\underline{u}}Z^{i}\, ,
\hspace{1cm}
e^{U}B^{i}\; =\;  \partial_{\underline{z}}Z^{i}\, ,\,\,\, \Rightarrow 
\partial_{\underline{z}^{*}} (e^{U}B^{i})\; =\; 0\, .
\end{equation}

Finally, the most general form that $\hat{n}$ can take in this case is

\begin{equation}
\hat{n}= dv + H du +\hat{\omega}\, ,   
\hspace{1cm}
\hat{\omega}=\omega_{\underline{z}}dz +\omega_{\underline{z}^{*}}dz^{*}\, ,
\end{equation}

\noindent
where all the functions in the metric are independent of $v$.
The above form of the null
tetrad components leads to a Brinkmann pp-wave metric \cite{kn:Br1}\footnote{
  The components of the
  connection and the Ricci tensor of this metric can be found in the
  Appendix of Ref.~\cite{Bellorin:2005zc}.
}

\begin{equation}
\label{eq:Brinkmetric}
ds^{2} \; =\; 2 du\ (dv \ +\ H du \ +\ \hat{\omega})
       \; -\; 2e^{2U}\ dzdz^{*}\, .
\end{equation}

As we now have a coordinate representation at our disposal, we can start checking
out the consistency conditions in this representation:
Let us expand the connection $\zeta$ as
\begin{equation}
\zeta \; =\; i\zeta_{n}\ \hat{n} 
      \ +\ i\zeta_{l}\ \hat{l}
      \ +\ \zeta_{m} \hat{m}
      \ -\ \zeta_{m^{*}}\hat{m}^{*} \; ,
\end{equation}
where $\zeta_{l}$ and $\zeta_{n}$ are real functions, whereas $\zeta_{m}$ is complex. 
Likewise expand
\begin{equation}
  \hat{\mathfrak{a}}\; =\; \mathfrak{a}_{l}\ \hat{l}\ +\ \mathfrak{a}_{m}\ \hat{m}
                     \ +\ \mathfrak{a}_{m^{*}}\ \hat{m}^{*}
                     \ +\ \mathfrak{a}_{n}\ \hat{n} \; ,
\end{equation}
and
\begin{equation}
\mathcal{Q}\; =\; \mathcal{Q}_{l}\ \hat{l} 
           \ +\ \mathcal{Q}_{m}\ \hat{m}\ +\ \mathcal{Q}_{m^{*}}\ \hat{m}^{*}
           \ +\ \mathcal{Q}_{n}\ \hat{n},
\end{equation}
where, due to the reality of $\mathcal{Q}$, $(\mathcal{Q}_{m})^{*}=\mathcal{Q}_{m^{*}}$.
Let us now consider the tetrad integrability equations
(\ref{eq:dtetrad1})-(\ref{eq:dtetrad3}): Eq. (\ref{eq:dtetrad1}) is 
by construction identically satisfied. 
Eq. (\ref{eq:dtetrad3}), with our choice of coordinate $z$ Eq.~(\ref{eq:z}),
implies

\begin{eqnarray}
 0 & =& e^{-U}\partial_{\underline z^{*}}U\ +\
        2\zeta_{m^{*}}\ -\ i\mathcal{Q}_{m^{*}} \; ,\label{eq:mstar}\\
 0 & =& -2i\zeta_{n}\ -\ i\mathcal{Q}_{n} \; ,
\end{eqnarray}
and 
\begin{equation}
\hat{\mathfrak{a}} \; = \; \left[
   \dot{U} 
   \ -\ 2i\zeta_l
   \ -\ i\mathcal{Q}_{l}
\right]\ \hat{m}  \; +\; \mathfrak{a}_{l}\ \hat{l}\, ,\label{eq:kuzeta}
\end{equation}

\noindent
where $\mathfrak{a}_{l}=\mathfrak{a}_{l}(z,z^{*},u)$ 
is a functions to be determined and dots
indicate partial derivation {\em w.r.t.\/} the coordinate $u$.  
Eq.~(\ref{eq:ziuz}) implies that $\zeta_{n}=\mathcal{Q}_{n}=0$ and from 
Eq.~(\ref{eq:mstar}) we obtain
\begin{equation}
\partial_{\underline z^*}(U+{\textstyle\frac{1}{2}}\mathcal K)=-2\zeta_{\underline z^*}.
\end{equation}
This last equation states that $\zeta_m^*$, whence also $\zeta_m$, can be eliminated by a 
gauge transformation, after which we are left with
\begin{equation}\label{eq:zetal}
\hat{\zeta} \; =\; i\zeta_{l}\ \hat{l}\; .
\end{equation}

At this point it is wise to return to Eq.~(\ref{eq:dzetamstar}) and
to deduce
\begin{eqnarray}
\mathsf{H}_{uv}\ \partial_{\mu} q^{u}\;  m^{*\nu}\partial_{\nu} q^{v}\; =\; 
(d\zeta)_{\mu\nu}\ m^{*\ \nu} & =&
   2e^{-U}(\partial_{\underline z}\zeta_{l}\ m_{[\mu}\ l_{\nu]}
   \ +\ \partial_{\underline z^{*}}\zeta_{l}\ m^{*}_{[\mu}l_{\nu]})m^{*\nu} \nonumber \\
 & & \nonumber \\
& =& e^{-U}\partial_{\underline z^{*}}\zeta_{l}\ l_{\mu} \; .
\end{eqnarray}
\par
This equation implies that $dq^{u}\sim \hat{l}$, and we are therefore obliged 
to accept the fact that in the null case, the hyperscalars can only depend
on the spacetime coordinate $u$!
\par
Had we been hoping for the hyperscalars to exhibit some interesting 
spacetime dependency, then this result would have been a bit of an anti-climax.
But then, the fact that the hyperscalars
can only depend on $u$, means that we can eliminate the connection $\mathsf{A}$ from the 
initial set-up, which means that as far as solutions to the Killing Spinor
equations is concerned, the problem splits into two disjoint parts: 
one is the solution to the KSEs in the null case of $N=2$ $d=4$ supergravity,
which are to be found in \cite{Tod:1983pm,Meessen:2006tu}, and the solutions
to Eq. (\ref{eq:hypersusyrule}).
\par
In the case at hand Eq.~(\ref{eq:hypersusyrule}) reduces to 
\begin{equation}
  0 \; =\; \mathsf{U}^{\alpha I}_{v}\varepsilon_{IJ}\; 
           \partial_{u}q^{v}\, \gamma^{u}\epsilon^{J}\; ,
\end{equation}
so that either we take the hyperscalars to be constant or impose the condition
$\gamma^{u}\epsilon^{I} =0$. This last condition is however always satisfied
by any non-maximally supersymmetric solution of the null case, to wit
Minkowski space and the 4D Kowalski-Glikman wave.  It is however obvious that
these solutions are incompatible with $u$-dependent hyperscalars, and its
reason takes us to the last point in this exposition: the equations of motion.
\par
As far as the equations of motion are concerned, it is clear that, since we
are dealing with a pp-wave metric, the hyperscalar equation of motion is
identically satisfied.  As the only coupling between vector multiplets and
hypermultiplets is through the gravitational interaction, see
Eq.~(\ref{eq:EOMGrav}), the only equation of motion that changes is the one in
the $uu$-direction.  More to the point, its sole effect is to change the
differential equation \cite[(5.91)]{Meessen:2006tu} determining the wave
profile $H$ in (\ref{eq:Brinkmetric}).
\par
A fitting example of a solution demonstrating just this, consider
the deformation of the cosmic string (\ref{eq:CosmString}):
\begin{equation}
  \label{eq:HypCosmString}
  \begin{array}{rclrcl}
    ds^{2} & =& 2\ du\ \left(dv \ +\ \mathsf{H}(\dot{q},\dot{q})\ |z|^{2} \right) 
           \; -\; 2e^{-\mathcal{K}}\ dz\ dz^{*}\, ,\hspace{1cm}
    & Z^{i} & =& Z^{i}(z) \, , \\
      & & & &  & \\
    F^{\Lambda} & = &  0\, ,   & q^{w} & = & q^{w}(u) \, ,\\
  \end{array}
\end{equation}
which is a 1/2-BPS solution.


\section*{Acknowledgements}

T.O.~would like to thank M.M.~Fern\'andez for her long standing support.
This work has been supported in part by the Spanish Ministry of Science and
Education grant \textit{Gravitaci\'on y la teor\'{\i}a de supercuerdas}
BFM2003-01090, by the Comunidad de Madrid grant HEPHACOS P-ESP-00346, by the
{\em Fondo Social Europeo} through an I3P project and by the EU Research
Training Network \textit{Constituents, Fundamental Forces and Symmetries of
  the Universe} MRTN-CT-2004-005104.

\appendix

\section{Quaternionic K\"ahler Geometry}
\label{sec-QKG}

A \textit{quaternionic K\"ahler manifold} is a real $4m$-dimensional
Riemannian manifold $\mathsf{HM}$ endowed with a triplet of complex
structures $\mathsf{J}^{x}: T(\mathsf{HM})\rightarrow
T(\mathsf{HM})\, ,\,\,\, (x=1,2,3)$ that satisfy the quaternionic
algebra

\begin{equation}
\mathsf{J}^{x}  \mathsf{J}^{y} = -\delta^{xy} 
+\varepsilon^{xyz}\mathsf{J}^{z}\, ,    
\end{equation}

\noindent
and with respect to which the metric, denoted by $\mathsf{H}$, is Hermitean:

\begin{equation}
\mathsf{H}(\ \mathsf{J}^{x} X,\ \mathsf{J}^{x}Y\ )= \mathsf{H}(X,Y)\, , 
\hspace{1cm}
\forall X,Y \in  T(\mathsf{HM})\, .
\end{equation}

This implies the existence of a triplet of 2-forms
$\mathsf{K}^{x}(X,Y)\equiv \mathsf{H}(\ \mathsf{J}^{x}X ,Y)$
globally known as the $\mathfrak{su}(2)$-valued \textit{hyperK\"ahler 2-forms}.

The structure of quaternionic K\"ahler manifold requires an $SU(2)$
bundle to be constructed over $\mathsf{HM}$ with connection 1-form
$\mathsf{A}^{x}$ with respect to which the hyperK\"ahler 2-form is covariantly
closed, i.e.

\begin{equation}
\mathfrak{D}\mathsf{K}^{x}\; \equiv\; d\mathsf{K}^{x} 
+\varepsilon^{xyz}\ \mathsf{A}^{y}\wedge \mathsf{K}^{z}\; =\; 0\, .  
\end{equation}

\noindent
Then, depending on whether the curvature of this bundle 

\begin{equation}
\mathsf{F}^{x}\; \equiv\; d\mathsf{A}^{x} 
\ +\ {\textstyle\frac{1}{2}}\varepsilon^{xyz}\
     \mathsf{A}^{y} \wedge \mathsf{A}^{z}\, , 
\end{equation}

\noindent 
is zero or is proportional to the hyperK\"ahler 2-form

\begin{equation}
\mathsf{F}^{x}= \lambda \mathsf{K}^{x}\, ,
\hspace{1cm}
\lambda\in \mathbb{R}_{/\{0\}}\, ,
\end{equation}

\noindent
the manifold is a \textit{hyperK\"ahler manifold} or a
\textit{quaternionic K\"ahler manifold}, respectively.

The $SU(2)$ connection acts on objects with vectorial $SU(2)$ indices,
such as the chiral spinors in this article, as follows:

\begin{equation}
  \begin{array}{rcl}
\mathfrak{D} \xi_{I} & \equiv  & d\xi_{I} +\mathsf{A}_{I}{}^{J}\xi_{J}\, ,\\
& & \\
\mathfrak{D} \chi^{I} & \equiv  & d\chi^{I} +\mathsf{A}^{I}{}_{J}\chi^{J}\, .\\
\end{array}
\end{equation}

Consistency with the raising and lowering of vector $SU(2)$ indices
via complex conjugation requires

\begin{equation}
\mathsf{A}^{I}{}_{J} = (\mathsf{A}_{I}{}^{J})^{*}\, . 
\end{equation}

\noindent
If we, following Ref.~\cite{Andrianopoli:1996cm}, put

\begin{equation}
\mathsf{A}_{I}{}^{J}\; \equiv\; 
{\textstyle\frac{i}{2}}\ \mathsf{A}^{x}\ (\sigma_{x})_{I}{}^{J}\, ,  
\end{equation}

\noindent
we get

\begin{equation}
\mathsf{A}^{I}{}_{J} \ =\ 
{\textstyle\frac{i}{2}}\ \mathsf{A}^{x}\
(\varepsilon\sigma_{x}\varepsilon^{-1})^{I}{}_{J}
= -{\textstyle\frac{i}{2}}\ \mathsf{A}^{x}
\varepsilon^{IK}\ (\sigma_{x})_{K}{}^{L}\ \varepsilon_{LJ}\, .
\end{equation}

Consistency between the above definitions of $SU(2)$-covariant
derivatives, $\mathsf{A}_{I}{}^{J}$ and $SU(2)$ curvature\footnote{Of
  course, $\mathsf{F}_{I}{}^{J} \equiv
  {\textstyle\frac{i}{2}}\ \mathsf{F}^{x}\ (\sigma_{x})_{I}{}^{J}$.}
$\mathsf{F}^{x}$ requires that the 3 matrices $(\sigma_{x})_{I}{}^{J}$
satisfy

\begin{equation}
[\ \sigma_{x}\ ,\ \sigma_{y}\ ]_{I}{}^{J} \; =\; 
      -2i\varepsilon_{xyz}\ (\sigma_{z})_{I}{}^{J}\, ,
\end{equation}

\noindent
whence we can take them to be the (Hermitean, traceless) Pauli
matrices satisfying

\begin{equation}
\label{eq:Paulimatrices}
(\sigma_{x}\sigma_{y})_{I}{}^{J} \; =\; \delta_{xy}\ \delta_{I}{}^{J} 
\, -\, i\varepsilon_{xyz}\ (\sigma_{z})_{I}{}^{J}\, .
\end{equation}

It is convenient to use a Vielbein on $\mathsf{HM}$ having as ``flat''
indices a pair $\alpha I$ consisting of one $SU(2)$-index $I$ and one
$Sp(m)$-index $\alpha = 1,\cdots,2m$

\begin{equation}
\mathsf{U}^{\alpha I} \; =\; \mathsf{U}^{\alpha I}{}_{u}\ dq^{u}\, ,
\end{equation}

\noindent
where $u\ =\ 1,\ldots ,4m$ and from now on we shall refer to this object as the 
Quadbein. This Quadbein is related to the metric 
$\mathsf{H}_{uv}$ by

\begin{equation}
\mathsf{H}_{uv} \; =\; \mathsf{U}^{\alpha I}{}_{u}\ \mathsf{U}^{\beta J}{}_{v}\
\varepsilon_{IJ}\mathbb{C}_{\alpha\beta}\, ,
\end{equation}

\noindent
and, further, it is required that

\begin{equation}
  \begin{array}{rcl}
2\ \mathsf{U}^{\alpha I}{}_{(u}\ \mathsf{U}^{\beta J}{}_{v)}\ 
\mathbb{C}_{\alpha\beta} & = & \mathsf{H}_{uv} \varepsilon^{IJ}\, ,\\
& & \\
2m\ \mathsf{U}^{\alpha I}{}_{(u}\ \mathsf{U}^{\beta J}{}_{v)}\ 
\varepsilon_{IJ} & = & \mathsf{H}_{uv}\ \mathbb{C}^{\alpha}\, ,\\
& & \\
\mathsf{U}_{\alpha I\, u} & \equiv & (\mathsf{U}^{\alpha I}{}_{u})^{*}
\; =\; \varepsilon_{IJ}\mathbb{C}_{\alpha\beta}\ \mathsf{U}^{\beta J}{}_{u}\, .
\end{array}
\end{equation}

\noindent
The inverse Quadbein $\mathsf{U}^{u}{}_{\alpha I}$ satisfies

\begin{equation}
\mathsf{U}_{\alpha I}{}^{u}\ \mathsf{U}^{\alpha I}{}_{v} =\delta^{u}{}_{v}\, , 
\end{equation}

\noindent
and, therefore, 

\begin{equation}
\label{eq:QuadbeinConst}
\mathsf{U}_{\alpha I}{}^{u} = \mathsf{H}^{uv}\ \varepsilon_{IJ}\mathbb{C}_{\alpha\beta}\
\mathsf{U}^{\beta J}{}_{v}\, .
\end{equation}

The Quadbein satisfies a \textit{Vielbein postulate},
{\em i.e.\/} they are covariantly constant with respect to the standard
Levi-Civit\`a connection $\Gamma_{uv}{}^{w}$, the $SU(2)$ connection
$\mathsf{A}_{u\, I}{}^{J}$ and the $Sp(m)$ connection
$\Delta_{u}{}^{\alpha\beta}$:

\begin{equation}\label{eq:FullWax}
\mathsf{D}_{u}\ \mathsf{U}^{\alpha I}{}_{v}
\; =\; \partial_{u}\mathsf{U}^{\alpha I}{}_{v}
\ -\ \Gamma_{uv}{}^{w}\ \mathsf{U}^{\alpha I}{}_{w} 
\ +\ \mathsf{A}_{u}{}^{I}{}_{J}\ \mathsf{U}^{\alpha J}{}_{v}
\ +\ \Delta_{u}{}^{\alpha\beta}\ \mathsf{U}^{\gamma I}{}_{v}
\mathbb{C}_{\beta\gamma}=0\, .
\end{equation}

\noindent
This postulate relates the three connections and the respective curvatures,
leading to the statement that the holonomy of a quaternionic K\"ahler manifold
is contained in $Sp(1)\cdot Sp(m)$, {\em i.e.\/}

\begin{equation}
R_{ts}{}^{uv}\ \mathsf{U}^{\alpha I}{}_{u}\ \mathsf{U}^{\beta J}{}_{v} 
\, +\, \varepsilon^{IK}\ \mathsf{F}_{ts\, K}{}^{J}\ \mathbb{C}^{\alpha\beta}
\, -\, 2\ \overline{R}_{ts}{}^{\alpha\beta} \varepsilon^{IJ}\; =\; 0\, ,
\end{equation}

\noindent where

\begin{equation}
\overline{R}_{ts}{}^{\alpha\beta} \; =\; 2\partial_{[t}\Delta_{s]}{}^{\alpha\beta}  
+2\Delta_{[t}{}^{\alpha\gamma}\ \Delta_{s]}{}^{\delta\beta}\mathbb{C}_{\gamma_{\delta}}\, ,
\end{equation}

\noindent is the curvature of the $Sp(m)$ connection.

A useful relation is 
\begin{equation}
\label{eq:su2curvaturevsquaternionivielbeins}
\mathsf{F}_{\mu\nu\, I}{}^{I}=2\lambda \mathsf{U}_{uI\alpha}  
\mathsf{U}_{v}{}^{J\alpha}\partial_{[\mu}q^{u} \partial_{\nu]}q^{v}\, .
\end{equation}

\section{C-map and dual quaternionic manifolds}
\label{appsec:dualquat}
The c-map is a manifestation of the T-duality between
the type IIA and IIB theories, compactified on the same
Calabi-Yau 3-fold. Since T-duality in supergravity theories 
is implemented by dimensional reduction, 
to be told that the c-map is derived by dimensionally
reducing an $N=2$ $d=4$ SUGRA coupled to $n$ vector- and $m$
hypermultiplets to $d=3$, and dualizing every vector field into a 
scalar field, should not come as too big a surprise. 
\par
In order to derive the c-map, consider the, rather standard, KK-Ansatz
\begin{equation}
  \label{eq:CMapAnsatz}
  \begin{array}{lclclcl}
    \hat{e}^{a} & =& e^{-\phi}\ e^{a} &\hspace{.5cm};\hspace{.5cm}&
    \hat{e}^{\underline{y}} & =& e^{\phi}\ (dy +A) \; ,\\
    & & & & & & \\
    \hat{A}^{\Lambda} & =& B^{\Lambda} +C^{\Lambda}\ (dy +A) & \rightarrow&
    \hat{F}^{\Lambda} & =& F^{\Lambda} \ +dC^{\Lambda}\wedge (dy +A) \; ,\\
    & & & &  & &  \\
    F^{\Lambda} & =& dB^{\Lambda} \ +\ C^{\Lambda}\ F & ,& F & =& dA \; ,
  \end{array}
\end{equation}
and use it on the action (\ref{eq:action}); the resulting action reads
\begin{eqnarray}
  \label{eq:CMapActionPre}
  \mathcal{S}_{(3)} & =& \int d^{3}\sqrt{g}\left[
           \textstyle{\frac{1}{2}}R 
           \ + d\phi^{2} 
           \ - e^{-2\phi} \mathrm{Im}(\mathcal{N})_{\Lambda\Sigma}\ dC^{\Lambda}dC^{\Sigma}
           \ +\mathcal{G}_{ij^{*}}dZ^{i}d(Z^{j})^{*} 
           \ + \mathsf{H}_{uv} dq^{u}dq^{v} 
  \right] \nonumber \\
  & & \nonumber \\
  & & +\int_{3}\left(\,
         \textstyle{\frac{1}{2}} \mathfrak{F}^{T}\ M\wedge\ *\mathfrak{F}
         \ +\ 
         \mathfrak{F}^{T}\ \wedge\ Qd\mathfrak{C} 
      \right) \; ,
\end{eqnarray}
where we have defined the $(\overline{n}+1)$-vectors 
$\mathfrak{F}^{T}\ =\ (dB^{\Lambda} ,dA )$ and $\mathfrak{C}^{T} = (C^{\Lambda},0)$.
Furthermore the $(\overline{n}+1)\times (\overline{n}+1)$-matrices $M$ and $Q$ are given
by
{\small
\begin{equation}
  M \ =\ 2e^{2\phi}\ \left(
    \begin{array}{lcl}
      \mathrm{Im}(\mathcal{N}) &~& \mathrm{Im}(\mathcal{N})\cdot C \\
       & & \\
      C^{T}\cdot\mathrm{Im}(\mathcal{N}) & &  
       C^{T}\cdot\mathrm{Im}(\mathcal{N})\cdot C -\textstyle{\frac{e^{2\phi}}{4}}
    \end{array}
  \right) \; ; \;
  Q \ =\ 2\left(
    \begin{array}{lcl}
      \mathrm{Re}(\mathcal{N}) &~& 0 \\
       & & \\
      C^{T}\cdot\mathrm{Re}(\mathcal{N}) & & 0
    \end{array}
  \right) \; .
\end{equation}
}
The field strengths can then be integrated out by adding to the above action
a Lagrange multiplier term $\mathfrak{F}^{T}\wedge\ d\mathfrak{L}$, imposing
the Bianchi identity $d\mathfrak{F}=0$. $\mathfrak{F}$ can then be integrated 
out by using its equation of motion 
$*\mathfrak{F} = M^{-1}(d\mathfrak{L}\ +Q\ d\mathfrak{C})$, resulting in
3d gravity coupled to a sigma model describing two disconnected quaternionic
manifolds, one with metric $\mathsf{H}_{uv}dq^{u}dq^{v}$, and the other one coming from
the gravity- and vector multiplets. Taking $\mathfrak{L}^{T}=(T_{\Lambda},\theta )$
we can write the metric of this $4\overline{n}$-dimensional quaternionic manifold as
{\small\begin{eqnarray}
  \label{eq:CMapDualMet}
  ds^{2}_{DQ} & =& d\phi^{2} 
    -e^{-2\phi}\mathrm{Im}(\mathcal{N})_{\Lambda\Sigma}\ dC^{\Lambda}dC^{\Sigma}
    \ + e^{-4\phi}\left( d\theta -C^{\Lambda}dT_{\Lambda}\right)^{2} 
    \ + \mathcal{G}_{ij}^{*}\ dZ^{i}\ d(Z^{j})^{*} \nonumber \\
    & & \nonumber \\
    & &  -\textstyle{\frac{1}{4}} e^{-2\phi}\mathrm{Im}(\mathcal{N})^{-1|\Lambda\Sigma}\
      \left( 
       dT_{\Lambda}+2\mathrm{Re}(\mathcal{N})_{\Lambda\overline{\Lambda}}dC^{\overline{\Lambda}}
      \right)\left(
       dT_{\Sigma}+2\mathrm{Re}(\mathcal{N})_{\Sigma\overline{\Sigma}}dC^{\overline{\Sigma}}
      \right) .
\end{eqnarray}}
The fact that this metric is indeed quaternionic was proven in \cite{Ferrara:1989ik}.
This kind of quaternionic manifolds is, for an obvious reason, called 
{\em dual quaternionic manifolds}, and is generically characterized
by the existence of at least $2(\overline{n}+1)$-translational isometries
\cite{deWit:1990na}, generated
by the following Killing vectors
\begin{equation}\begin{array}{lclclcl}
U & =& \partial_{\phi} +T_{\Lambda}\partial_{\scriptstyle{T_{\Lambda}}}
+C^{\Lambda}\partial_{\scriptstyle{C^{\Lambda}}} +2\theta\partial_{\theta} &\hspace{.1cm};\hspace{.1cm}& V & =& \partial_{\theta} \; , \\
   & & & & & & \\
X^{\Lambda} & =& \partial_{\scriptstyle{T_{\Lambda}}} & ,&
Y_{\Lambda} & =& \partial_{\scriptstyle{C^{\Lambda}}} 
                +T_{\Lambda}\partial_{\theta} \; .
\end{array}\end{equation}
These vector fields satisfy the commutation relation of a Heisenberg
algebra, {\em i.e.}
\begin{equation}
  \label{eq:CMapHeis}
  \begin{array}{lclclcl}
    \left[ U, X^{\Lambda} \right] & =& -X^{\Lambda} &\hspace{.2cm},\hspace{.2cm}&
    \left[ U, Y_{\Lambda} \right] & =& -Y_{\Lambda} \; ,{}\\
     & & & &  & & {}\\
    \left[ U, V \right] & =& -2\ V & ,&  
    \left[ X^{\Lambda},Y_{\Sigma} \right] & =& \delta^{\Lambda}{}_{\Sigma}\ V \; . 
  \end{array}
\end{equation}
The automorphism group of this Heisenberg algebra is $Sp(\overline{n},\mathbb{R})$, and
one can find a nice $Sp(\overline{n},\mathbb{R})$-adapted coordinate system by doing 
the coordinate transformation
$T_{\Lambda}\rightarrow -2T_{\Lambda}$ and $\theta \rightarrow \theta -C^{\Lambda}T_{\Lambda}$;
this transformation allows us to write the metric, introducing the real symplectic vector 
$\mathcal{S}^{T} = (C^{\Lambda},T_{\Lambda})$, as
\begin{equation}\label{eq:CMapMetFinal}
  ds^{2}_{DQ} \ =\ d\phi^{2} 
       + \mathcal{G}_{ij^{*}} dZ^{i}d(Z^{j})^{*}
       + e^{-4\phi}\left( d\theta -\langle\mathcal{S}|d\mathcal{S}\rangle\right)^{2}
       + d\mathcal{S}^{T}\ \mathfrak{M}\ d\mathcal{S} \; ,
\end{equation}
where $\mathfrak{M}$ is the $2\overline{n}\times 2\overline{n}$-matrix
\begin{eqnarray}
  \label{eq:CMapMMatrix}
   \mathfrak{M} & =& -\left(
    \begin{array}{ccc}
      \mathrm{Im}(\mathcal{N}) \ +\ 
        \mathrm{Re}(\mathcal{N})\mathrm{Im}(\mathcal{N})^{-1}\mathrm{Re}(\mathcal{N})
      & ~& -\mathrm{Re}(\mathcal{N})\mathrm{Im}(\mathcal{N})^{-1} \\
      & & \\
      -\mathrm{Im}(\mathcal{N})^{-1}\mathrm{Re}(\mathcal{N}) & &
      \mathrm{Im}(\mathcal{N})^{-1}
    \end{array}
   \right) \; , \nonumber \\
   & & \nonumber \\
     & =& 2\Omega\ \mathrm{Re}\left( 
              \mathcal{V}\ \mathcal{V}^{\dagger} \ +\ 
              \mathcal{U}_{i}\ \mathcal{G}^{ij^{*}}\ \mathcal{U}_{j}^{\dagger}
           \right)\Omega^{T} \; ,
\end{eqnarray}
where $\Omega$ is the inner product left invariant by $Sp(\overline{n};\mathbb{R})$.
Moreover, $\mathfrak{M}$ is positive definite and has the correct and obvious properties \cite{Ferrara:1996dd}
to make the metric $Sp(\overline{n},\mathbb{R})$-covariant.
\par
In order to discuss the Quadbein, it is convenient to split the 
$\alpha=1\ldots 2\overline{n}$
index as $\alpha\ \rightarrow\ (\Lambda \bar{\alpha})$ where the new 
$\bar{\alpha} =1,2$ and 
as usual $\Lambda = 0,\ldots ,n$. This means that we split 
$\mathbb{C}_{\alpha\beta} \ =\ \delta_{\Lambda\Sigma}\ \varepsilon_{\bar{\alpha}\bar{\beta}}$
and a base for the matrices satisfying Eq. (\ref{eq:QuadbeinConst}) can be found
with great ease, but since it will not be needed, we shall abstain from presenting
them here.
\par
It is likewise convenient to introduce the objects ($a=1,\ldots ,n_{V}$) 
\begin{equation}
  \label{eq:CMapDefs}
  \begin{array}{lclclcl}
    E^{\Lambda} & =& \left( E^{0},E^{a}\right) &\;\; ;\;\;&
    \sqrt{2}\ E^{0} & =& 
          d\phi +i\ e^{-2\phi}\left[ d\theta -\langle S\mid dS\rangle\right] \; ,\\[.2cm]
     & & & & E^{a}\overline{E^{a}} & =& 
          \textstyle{\frac{1}{2}} \mathcal{G}_{ij^{*}}dZ^{i}d(Z^{j})^{*} \\
     & & && && \\
     \mathcal{U}^{\Lambda} & =& \left( \mathcal{V},\mathcal{U}^{a}\right) & ;&
     \mathcal{U}^{a} & =& \mathcal{U}_{i}\ \mathcal{G}^{ij^{*}} \overline{E^{a}_{j}} \; .
     \end{array}
\end{equation}
With these definitions we can write the Quadbein compactly and manifestly
$Sp(\overline{n};\mathbb{R})$-covariant as
\begin{equation}
  \label{eq:CMapQuadbein}
  \mathsf{U}^{(\Lambda\bar{\alpha})\ I} \; =\; 
    \left(
      \begin{array}[lclclcl]{ccc}
         E^{\Lambda}  &    e^{-\phi}\ \langle dS\mid \overline{\mathcal{U}}^{\Lambda}\rangle \\
         -e^{-\phi}\ \langle dS\mid \mathcal{U}^{\Lambda}\rangle 
                  & (E^{\Lambda})^{*}
      \end{array}
    \right) \; ,
\end{equation}

 In this parametrization, the $\mathfrak{sp}(1)$ connection can be seen to be
 \begin{equation}
   \label{eq:CMapSP1Con}
   \mathsf{A}_{I}{}^{J} \; =\; \textstyle{\frac{i}{2}}\left(
     \begin{array}{ccc}
        \mathcal{Q} -\sqrt{2}\mathrm{Im}( E^{0} ) &\hspace{.3cm}~& 
          -2\sqrt{2}i\ e^{-\phi}\langle dS\mid \overline{\mathcal{V}}\rangle \\
        & & \\
        2\sqrt{2}i\ e^{-\phi}\langle dS\mid \mathcal{V}\rangle & & 
        \sqrt{2}\mathrm{Im}( E^{0} ) -\mathcal{Q}   
     \end{array}
   \right) \; .
 \end{equation}
 Let us close this appendix with some comments: An interesting quaternionic
 manifold is the so-called universal quaternionic manifold, which is the
 manifold that arises from applying the c-map on minimal $N=2$ $d=4$ SUGRA:
 it is therefore given by the formulae in this section for $n=0$. From the
 parent discussion it is then also paramount that we are dealing with a
 homogeneous space; It is admittedly less paramount that the universal
 quaternionic manifold is the symmetric space $SU(1,2)/U(2)$, but a quite
 standard calculation shows this to be the case.
\par
We derived the c-map through dimensional reduction over a spacelike circle.
Similarly one can dimensionally reduce the action over a timelike circle,
resulting in a space of signature $(2\overline{n},2\overline{n})$ and whose
holonomy is contained in $Sp(1,\mathbb{R})\cdot Sp(\overline{n})$.  In the
rigid limit, {\em i.e.} when $\lambda =0$, one recovers the
$(1,2)$/para-hyperK\"ahler structure discussed in {\em e.g.\/}
\cite{thesis:kath,Cortes:2005uq} The para-universal para-quaternionic
manifold, {\em i.e.} the manifold one obtains by the timelike c-map from
minimal $N=2$ $d=4$ SUGRA, can be seen to be $SU(1,2)/U(1,1)$.


\begin{thebibliography}{99}

\bibitem{Meessen:2006tu}
P.~Meessen and T.~Ort\'{\i}n,
``{\em The supersymmetric configurations of N = 2, d = 4 
supergravity coupled to vector supermultiplets}'', 
to be published in Nucl. Phys \textbf{B}.
[arXiv:\hepth{0603099}].

\bibitem{deWit:1984pk}
B. de Wit and A. van Proeyen,
Nucl.\ Phys.\ B {\bf 245} (1984) 89.

\bibitem{deWit:1984px}
B. de Wit, P.G. Lauwers and A. van Proeyen,
Nucl.\ Phys.\ B {\bf 255} (1985) 569.

\bibitem{Celi:2003qk}
A.~Celi,
``{\em Toward the classification of BPS solutions of N = 2, d = 5 gauged
       supergravity with matter couplings}'',
arXiv:\hepth{0405283}.

\bibitem{Tod:1983pm}
K.P. Tod,
Phys.\ Lett.\ B {\bf 121} (1983) 241;
Class. Quant. Grav. {\bf 12} (1995), 1801.

\bibitem{kn:Wilczek}
F. Wilczek, in: \textit{Quark confinement and field
theory}, ed. Stump and Weingarte (Wiley-Interscience, New York,
1977)

\bibitem{Charap:1977ww}
J.M.~Charap and M.J.~Duff,
Phys.\ Lett.\ B {\bf 69} (1977) 445.

\bibitem{Candelas:1985en}
P. Candelas, G.T. Horowitz, A. Strominger and E. Witten,
Nucl.\ Phys.\ B {\bf 258} (1985) 46.

\bibitem{Khuri:1992hk}
R.R. Khuri,
Nucl.\ Phys.\ B {\bf 387} (1992) 315
[arXiv:\hepth{9205081}].

\bibitem{Gauntlett:1992nn}
J.P.~Gauntlett, J.A.~Harvey and J.T.~Liu,
Nucl.\ Phys.\ B {\bf 409} (1993) 363
[arXiv:\hepth{9211056}].

\bibitem{Duff:1993yb}
M.J.~Duff, R.R.~Khuri, R.~Minasian and J.~Rahmfeld,
Nucl.\ Phys.\ B {\bf 418} (1994) 195
[arXiv:\hepth{9311120}].

\bibitem{Kallosh:1994wy}
R.~Kallosh and T.~Ort\'{\i}n,
Phys.\ Rev.\ D {\bf 50} (1994) 7123
[arXiv:\hepth{9409060}].

\bibitem{Maldacena:2000mw}
J.M.~Maldacena and C.~N\'u\~{n}ez,
Int.\ J.\ Mod.\ Phys.\ A {\bf 16} (2001) 822
[arXiv:\hepth{0007018}].


\bibitem{Denef:2000nb}
F.~Denef,
JHEP {\bf 0008} (2000) 050
[arXiv:\hepth{0005049}];
%
B.~Bates and F.~Denef,
``{\em Exact solutions for supersymmetric stationary black hole composites}'',
arXiv:\hepth{0304094}.

\bibitem{Bellorin:2006xr}
J.~Bellor\'{\i}n, P.~Meessen and T.~Ort\'{\i}n,
``{\em Supersymmetry, attractors and cosmic censorship}'',
arXiv:\hepth{0606201}.

\bibitem{Behrndt:1997ny}
K. Behrndt, D. L\"ust and W.A. Sabra,
Nucl.\ Phys.\ B {\bf 510} (1998) 264
[arXiv:\hepth{9705169}].

\bibitem{Andrianopoli:1996cm} 
L.~Andrianopoli, M.~Bertolini,
 A.~Ceresole, R.~D'Auria, S.~Ferrara, P.~Fr\'e and T.~Magri,
J.\ Geom.\ Phys.\  {\bf 23} (1997) 111
[arXiv:\hepth{9605032}].

\bibitem{Kallosh:1993wx}
R.~Kallosh and T.~Ort\'{\i}n,
``{\em Killing spinor identities}'',
arXiv:\hepth{9306085}.

\bibitem{Bellorin:2005hy}
J.~Bellor\'{\i}n and T.~Ort\'{\i}n,
Phys.\ Lett.\ B {\bf 616} (2005) 118
[arXiv:\hepth{0501246}].

\bibitem{Bellorin:2005zc}
J.~Bellor\'{\i}n and T.~Ort\'{\i}n,
Nucl.\ Phys.\ B {\bf 726} (2005) 171
[arXiv:\hepth{0506056}].

\bibitem{kn:Br1} H.W. Brinkmann,
                 {\it Proc.~Natl.~Acad.~Sci.~U.S.}~\textbf{9} (1923) 1;
                 {\it Math.~Annal.}~\textbf{94} (1925) 119.

\bibitem{Ferrara:1989ik}
S.~Ferrara and S.~Sabharwal,
Nucl.\ Phys.\ B {\bf 332} (1990) 317.

\bibitem{deWit:1990na}
B. de~Wit and A. van~Proeyen,
Phys.\ Lett.\ B {\bf 252} (1990) 221.

\bibitem{Ferrara:1996dd}
S. Ferrara and R. Kallosh,
Phys. Rev. D {\bf 54}(1996) 1514
[arXiv:\hepth{9602136}].

\bibitem{thesis:kath}
I.~Kath, 
``{\em Killing spinors on pseudo-Riemannian manifolds}'',
Habilitationsschrift an der Humboldt Universit\"at Berlin, 1999.

\bibitem{Cortes:2005uq}
V.~Cort\'es, C.~Mayer, T.~Mohaupt and F.~Saueressig,
JHEP {\bf 0506} (2005) 025 [arXiv:\hepth{0503094}].

\end{thebibliography}
\end{document}